\documentclass[aps,prl,twocolumn,superscriptaddress]{revtex4-2}

\usepackage{graphicx}% Include figure files
\usepackage{epsfig}
\usepackage{dcolumn}
\usepackage{bm}
\usepackage{overpic}
\usepackage{color}
\usepackage{multirow}
\usepackage{subfigure}
\usepackage{lineno}
\usepackage{amsmath}
\usepackage{url}
\usepackage[table]{xcolor}
\usepackage{diagbox}
\usepackage{multirow}
\usepackage{ulem}

\usepackage{xcolor}

\usepackage{hyperref}
\hypersetup{
colorlinks=true,
linkcolor=blue,
filecolor=blue,%magenta,
urlcolor=blue,%cyan,
citecolor=blue,
pdftitle={LaTeX Example},
pdfpagemode=FullScreen,
}

\begin{document}

%\linenumbers

%Title of paper
\title{A novel design of general-purpose spectrometer with nucleon polarimeter function}

\author{Chuang-Xin Lin}
\email[]{linchuangxin@impcas.ac.cn}
\affiliation{Institute of Modern Physics, Chinese Academy of Sciences, Lanzhou 730000, People's Republic of China}
\affiliation{University of Chinese Academy of Sciences, Beijing 100049, People's Republic of China}

\author{Xiao-Rong Lv}
\affiliation{Institute of Modern Physics, Chinese Academy of Sciences, Lanzhou 730000, People's Republic of China}

\author{Boxing Gou}
\email[]{gouboxing@impcas.ac.cn}
\affiliation{Institute of Modern Physics, Chinese Academy of Sciences, Lanzhou 730000, People's Republic of China}
\affiliation{University of Chinese Academy of Sciences, Beijing 100049, People's Republic of China}

\author{Ai-Qiang Guo}
\email[]{guoaq@impcas.ac.cn}
\affiliation{Institute of Modern Physics, Chinese Academy of Sciences, Lanzhou 730000, People's Republic of China}
\affiliation{University of Chinese Academy of Sciences, Beijing 100049, People's Republic of China}

\author{Yu-Tie Liang}
\email[]{liangyt@impcas.ac.cn}
\affiliation{Institute of Modern Physics, Chinese Academy of Sciences, Lanzhou 730000, People's Republic of China}
\affiliation{University of Chinese Academy of Sciences, Beijing 100049, People's Republic of China}

% \author{Nu Xu}
% %\email[]{}
% \affiliation{Institute of Modern Physics, Chinese Academy of Sciences, Lanzhou 730000, People's Republic of China}
% \affiliation{University of Chinese Academy of Sciences, Beijing 100049, People's Republic of China}

\date{\today}

\begin{abstract}
%The spin polarization of hadrons produced in particle and nuclear reactions is an extra observable associated with details of the underlying physics that cannot be obtained otherwise. In conventional experiments using a general-purpose spectrometer, the final-state polarization is not accessible. A novel technique for measuring the final-state nucleon polarization in a general-purpose spectrometer has been proposed. With the aim of implementing this function on the proposed hyperon-nucleon spectrometer, a systematic optimization has been performed. This first integration of a polarimeter into a general-purpose spectrometer provides a valuable benchmark for future experiments. By combining spin polarization data with the four-momentum of final-state particles, we can achieve a more comprehensive understanding of the physics in nuclear and particle physics research. 

The spin polarization of hadrons is a key observable for probing the details of particle and nuclear interactions, offering information not available from other measurements. Currently, general-purpose spectrometers lack the capability to access final-state polarization. A novel technique to measure nucleon polarization within such a spectrometer has been proposed. Using this technique, a new design based on the proposed hyperon-nucleon spectrometer is developed. Systematic optimization confirms that the nucleon polarimeter functions effectively without impairing the detector's conventional performance. This successful integration, the first attempt of the general-purpose spectrometer, sets a valuable benchmark for future experiments. Ultimately, correlating spin polarization with four-momentum data will lead to a more profound understanding of the underlying physics.

\end{abstract}

\maketitle

\section{Introduction}
Spin, being an intrinsic property of the microcosm, plays profound roles in the particle structure and dynamics.
However, in the most conventional and simplest experiments, where (unpolarized) cross sections or decay rates are measured, the contributions from different spin states are largely canceled out. 
To investigate the effects of spin upon how particles interact with each other and how they compose into more complex objects, one needs to perform spin-polarized experiments, which include initial- and final-state polarized experiments. The former involves preparation of polarized beams and/or targets, and the latter requires measurement of the spin polarization of final-state particles. 
Since the early 1960s, polarized targets and beams have been developed and employed at many laboratories, valuable experimental data on $NN$, $eN$ and $\gamma N$ processes have been accumulated~\cite{NNOnline,MAID_1998,*MAID_2007,*MAIDURL,SAID_2000,*SAID_2007,*SAIDURL}. These data have dramatically improved our understanding of nuclear force and nucleon structure. 

However, measurements of final-state polarization are much less. 
This is because the momentum and spin polarization of final-state particles are both angle-dependent, and it is usually difficult to find a reference process to calibrate polarization for particles with a broad energy distribution emitted at a wide angular range. 
An exception is when a particle in the exit channel decays through the weak interaction. It has long been recognized that the physical asymmetry of a weak process can be exactly calculated; thus weakly decayed hyperons like $\Lambda$ are spin-self-analyzable; that is, their polarization can be inferred from the angular distribution of the decay products, no extra reference reaction is needed. In addition, the spin alignment of vector meson, known as one kind of polarization, can be measured from its strong decay. 
In recent years, measurements of hyperon polarization or the spin alignment of vector meson have been made in $e^+ e^-$ and $AA$ collisions over a wide energy scale~\cite{BESIII:2018cnd,PhysRevLett.122.042001,PhysRevLett.36.1113, PhysRevLett.67.1193,STAR:2017ckg,STAR:2022fan, Chen:2024aom, PhysRevLett.94.102301, PhysRevC.77.044902, Liang:2004xn, Liu:2000fi, Zhangjinlong-NST, CHEN2023874, LU1995419, Xu:2005ru, Xu:2004es, xucao2024, Zhou:2008fb, PhysRevLett.134.022301}.

In contrast, the spin polarization of stable baryons, say a proton, can only be extracted when it participates in a spin-dependent process. This involves a secondary interaction, and is difficult to realize in a large acceptance general-purpose spectrometer. Over the past half century, dozens of large scale general-purpose spectrometers, such as BESIII~\cite{ABLIKIM2010345}, BelleII~\cite{ONUKI202278}, STAR~\cite{Chen:2024aom}, ATLAS~\cite{Aad_2024}, CMS~\cite{Hayrapetyan_2024}, etc., had been built and made tremendous contributions in the field of nuclear and particle physics. However, all these spectrometers are designed to measure the four momenta of stable final-state particles, without the capability to measure their spin polarization. 

Recently, a technique for measuring final-state nucleon polarization within a general-purpose spectrometer has been proposed~\cite{c642-1lzb}. Using this technique, we developed a new general-purpose spectrometer that achieves high-precision measurement of nucleon polarization without significantly compromising the conventional four-momenta measurements. In this paper, we present the novel design for a proposed hyperon-nucleon spectrometer at the High Intensity Heavy-Ion Accelerator Facility (HIAF)~\cite{YANG2013263}. A key innovation, which establishes a benchmark for future experiments, is the first integration of a polarimeter function within a general-purpose spectrometer. The paper details the detector optimization strategy, the methods for signal extraction and background suppression, and the estimation of the cross section and its associated uncertainties.

\section{General principle of nucleon polarimeter}

A measurement of the proton polarization usually utilizes the spin-dependent cross section for the $pp$ or $p\textrm{C}$ elastic scattering given their large analyzing powers. In the case of an unpolarized target, the differential cross section for a given center of mass polar angle $\theta$ is expressed as~\cite{PhysRevLett.90.142301, Bystricky:1976jr}: 
\begin{equation}
\frac{d\sigma}{d\phi d\!\cos\theta} =  \frac{1}{2\pi}\frac{d\sigma_0}{ d\!\cos\theta}\left[1+\mathcal{P}_yA_N\!(\theta)\cos\phi\right]
\label{eq:pol_XSC}
\end{equation}

\begin{figure}
\includegraphics[width=0.45\textwidth]{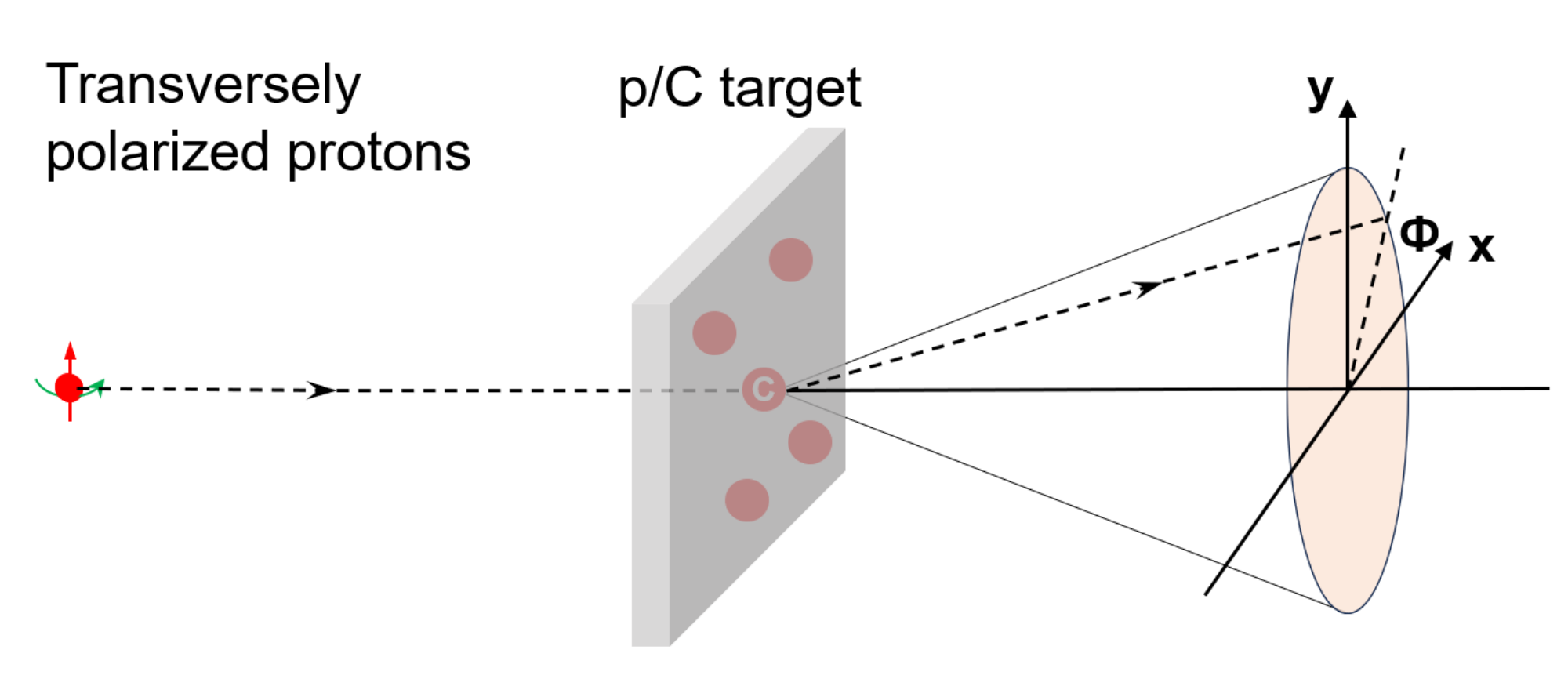}
\caption{\label{fig:pC_illustration} {Illustration of proton polarization measurement with $pp$ or $p\textrm{C}$ elastic scattering. 
}}
\end{figure}

Here, $\sigma_0$ represents the unpolarized cross section, $\theta$ and $\phi$ denote the polar and azimuthal angle of the scattered proton in the center of mass frame of scattering. $\mathcal{P}_y$ corresponds to the transverse proton polarization to be determined. $A_N(\theta)$ is the scattering angle-dependent analyzing power, which has been measured extensively~\cite{PhysRev.148.1289,PhysRev.163.1470, PhysRev.95.1348, ALBROW1970445, PhysRevC.24.1778, vonPrzewoski:1998ye, PhysRevLett.41.384, PhysRevD.21.580, PhysRevD.40.35, PhysRev.105.288, GREENIAUS1979308}.  Figure~\ref{fig:pC_illustration} shows the illustration of proton polarization measurement.

%~\cite{PhysRev.148.1289,PhysRev.163.1470, PhysRev.95.1348, ALBROW1970445}. 

%~\cite{PhysRev.148.1289,PhysRev.163.1470, PhysRev.95.1348, ALBROW1970445, PhysRevC.24.1778, vonPrzewoski:1998ye, PhysRevLett.41.384, PhysRevD.21.580, PhysRevD.40.35, PhysRev.105.288, GREENIAUS1979308}. 

Proton polarimeters utilizing $pp$ or $p\textrm{C}$ scattering has been widely employed, primarily as dedicated detectors for measuring initial-state proton beam polarization or final-state protons within limited acceptance, even though it is not yet implemented in large acceptance general-purpose spectrometer.

%Proton polarimeters utilizing $pp$ or $p\textrm{C}$ scattering has been widely employed, primarily as dedicated detectors for measuring initial-state proton beam polarization or final-state protons within limited acceptance. For a large acceptance general-purpose spectrometer, a novel technique has been developed to measure proton polarization based on the BESIII detector configuration~\cite{c642-1lzb}. 
%Using polarized proton from hyperon decay, this technique has been validated and proves that BESIII detector can work as a proton polarimeter. 

\section{H-NS at HIAF}

\begin{figure}
\includegraphics[width=0.5\textwidth]{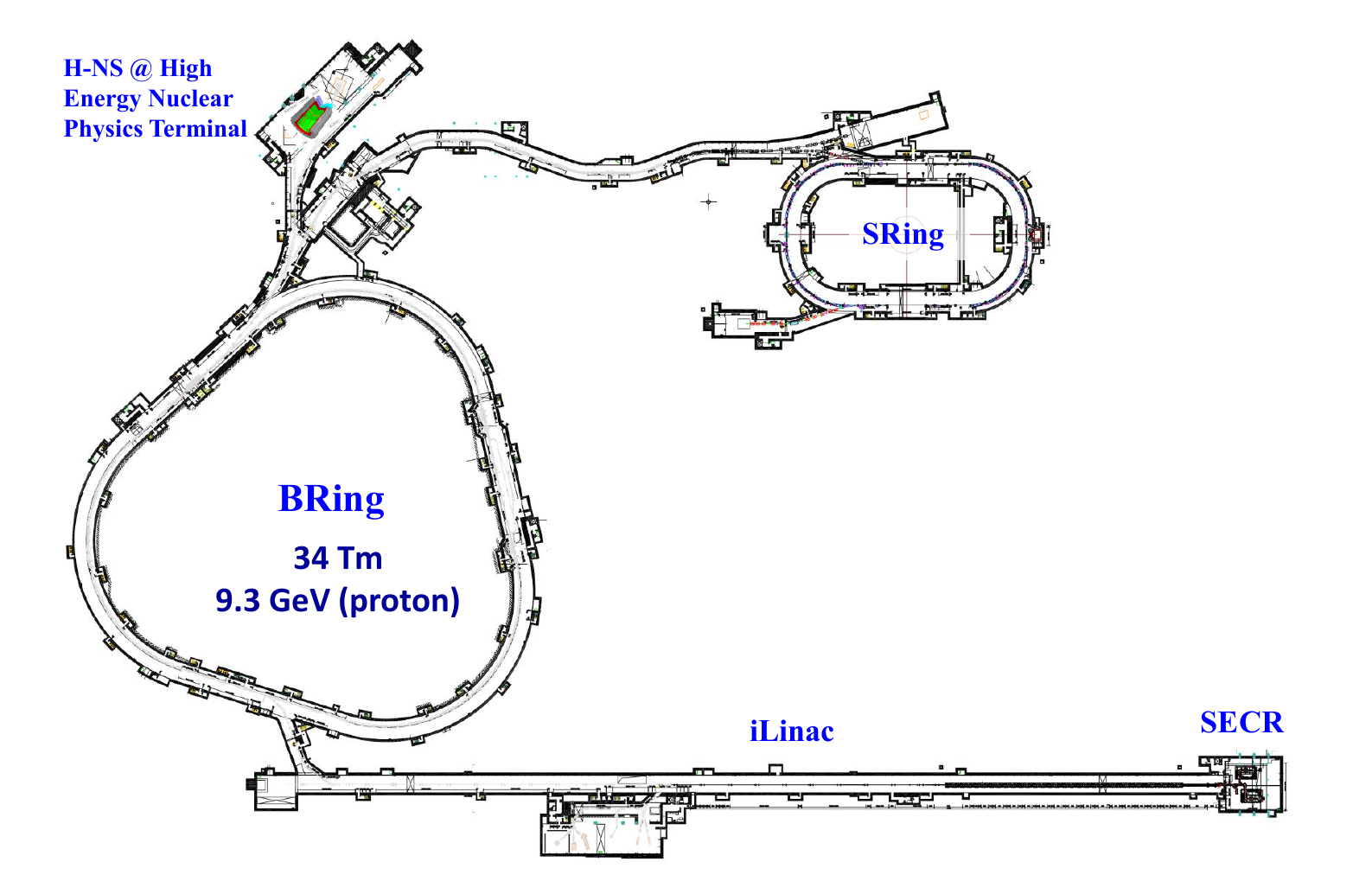}
\caption{\label{fig:HIAF} {Layout of the HIAF complex. H-NS is located in the high energy nuclear physcis terminal.
}}
\end{figure}

HIAF is a next-generation, internationally advanced heave-ion accelerator complex located in Huizhou, China. It is scheduled to begin operation by the end of 2025. HIAF is designed to produce ion beams—ranging from protons to uranium—across a wide range of energies at unprecedented intensities. The layout of the HIAF complex is shown in Fig.~\ref{fig:HIAF}. Ion beams generated by the superconducting ion source (SECR) are first accelerated to approximately 10 MeV per nucleon in the high-current superconducting linac (iLinac). These beams are then injected into the booster ring (BRing), which has a magnetic rigidity of 34 Tm, where they are further accelerated to energies of a few GeV per nucleon. For proton beams, the maximum energy reaches 9.3 GeV with an intensity of up to 2$\times10^{12}$ particles per pulse. A slow extraction system in the BRing can provide a quasi-continuous beam over several seconds for nuclear physics experiments in the high energy nuclear physics terminal.

%For proton beams, the maximum energy will reach up to 9.3 GeV, with plans to upgrade to 25 GeV in the facility’s second phase.

A proposed fix-target experiment at the high energy nuclear physics terminal of HIAF, called the Hyperon–Nucleon Spectrometer (H-NS), will utilize these high-intensity proton and ion beams to perform $pp/pA/AA$ collisions. Key research topics at H-NS include investigating the origin of hyperon polarization, probing the spin structure of baryons, conducting hadron spectroscopy, exploring the QCD phase diagram and searching for the critical point, etc.

\begin{figure}
\includegraphics[width=0.5\textwidth]{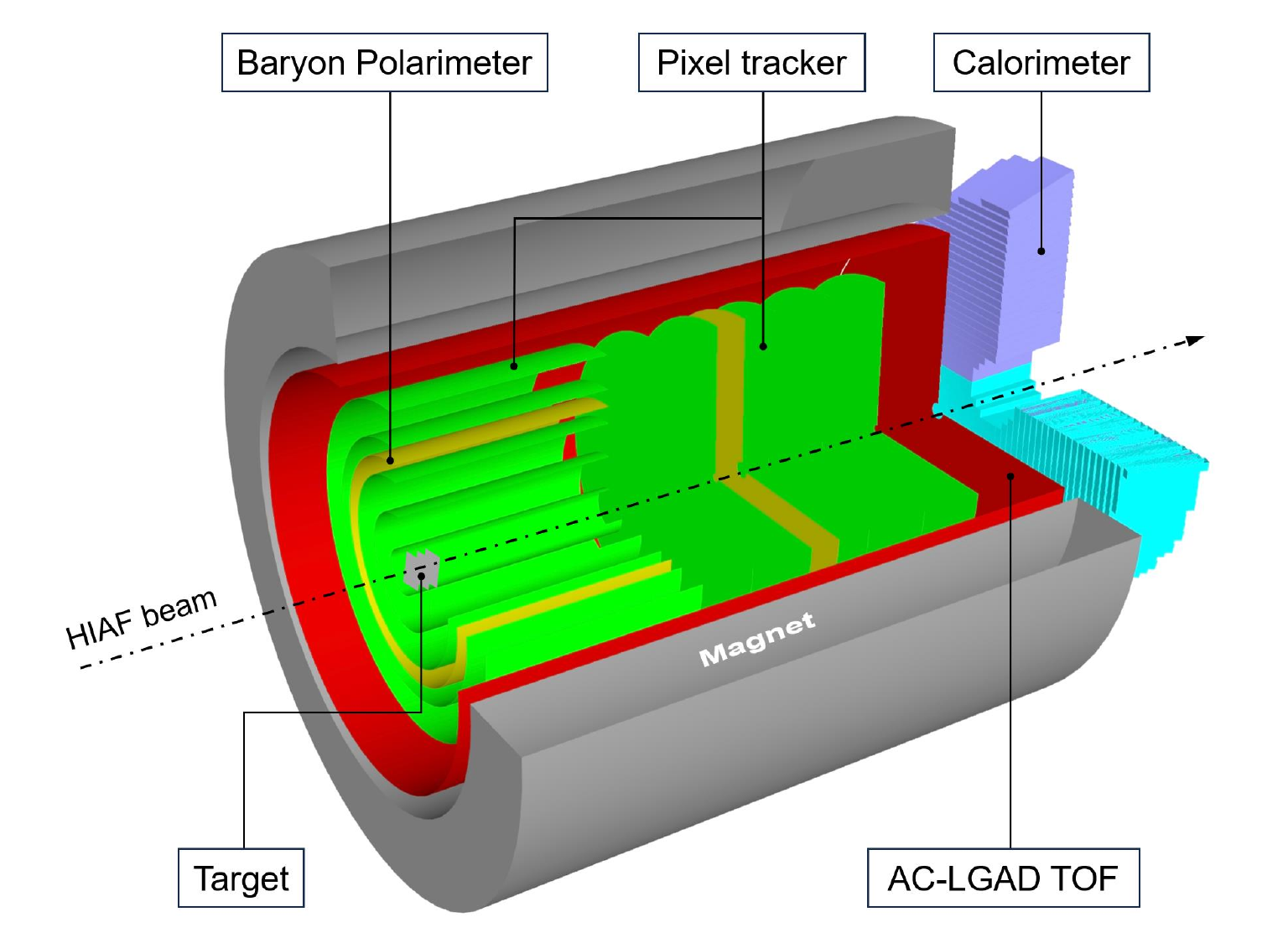}
\caption{\label{fig:HNS_det} {H-NS detector with proton polarimetry.
}}
\end{figure}

The H-NS detector is designed for high-precision measurement of hyperon and nucleon polarization under high-luminosity conditions. Its design demands excellent momentum and spatial resolution, as well as a fast response time. As shown in Fig.\ref{fig:HNS_det}, the conceptual design comprises three main subsystems: a pixel silicon tracker for charged particle tracking, a Low-Gain Avalanche Detector (LGAD)-based Time-of-Flight (TOF) system~\cite{Sadrozinski:2017qpv, Liukang-NST} for particle identification, and a electro-magnetic calorimeter (ECal)~\cite{ECal-NST} for neutral particle detection.
    
The tracking detector utilizes a modular Monolithic Active Pixel Sensor (MAPS)~\cite{RTurchetta_2006, Herui-NST, DuanFF-NST, Wangshen-NST, Cao-JINST, ZhaoC-JINST, Malong-NST}. The pixel size is 30\,$\mu$m, with a material budget of 0.35\% $\textrm{X/X}_0$ per layer. The tracker consists of a cylindrical barrel section and a disk-shaped forward section, each containing five detection layers. The barrel layers are evenly distributed with radii ranging from 5 cm to 37 cm, and the barrel length is approximately 50 cm. The forward section comprises five disks with positions evenly spaced along the beam axis from Z = 45 cm to Z = 105 cm. A carbon target layer, approximately 1 mm thick, is positioned between the third and fourth layers in both the barrel and forward sections to facilitate nucleon polarization measurements. This target contributes approximately 10\% of the total material budget in the tracking region.
    
The TOF detector is constructed from LGAD sensors, which provide time and spatial measurements. It achieves a time resolution of approximately 30 ps and a spatial resolution of about 200\,$\mu$m. The TOF system includes a cylindrical barrel and a forward disk. The barrel has a radius of 45 cm and a length of 150 cm. The forward disk is located at Z = 125 cm and has a radius of 45 cm. The TOF detector, with decent spatial resolution, will serve as an extra tracking layer and satisfy the minimum requirement for nucleon polarization measurement. The ECal detector is not used in the nucleon polarization measurement and will not be explained here. The entire detector is housed within a superconducting solenoid magnet that provides a magnetic field strength of 1.5 T.

%The calorimeter, positioned downstream of the forward TOF detector, is composed of approximately 1800 PbWO$_4$ crystals. It has a radius of 50 cm, with each crystal module having a length of about 20 cm. The calorimeter achieves an energy resolution of approximately $2\%/\sqrt{E \mathrm{(GeV)}}$ and a spatial resolution of about 2 mm.

\section{Impact on the conventional detector performance}

\begin{figure}
\includegraphics[width=0.5\textwidth]{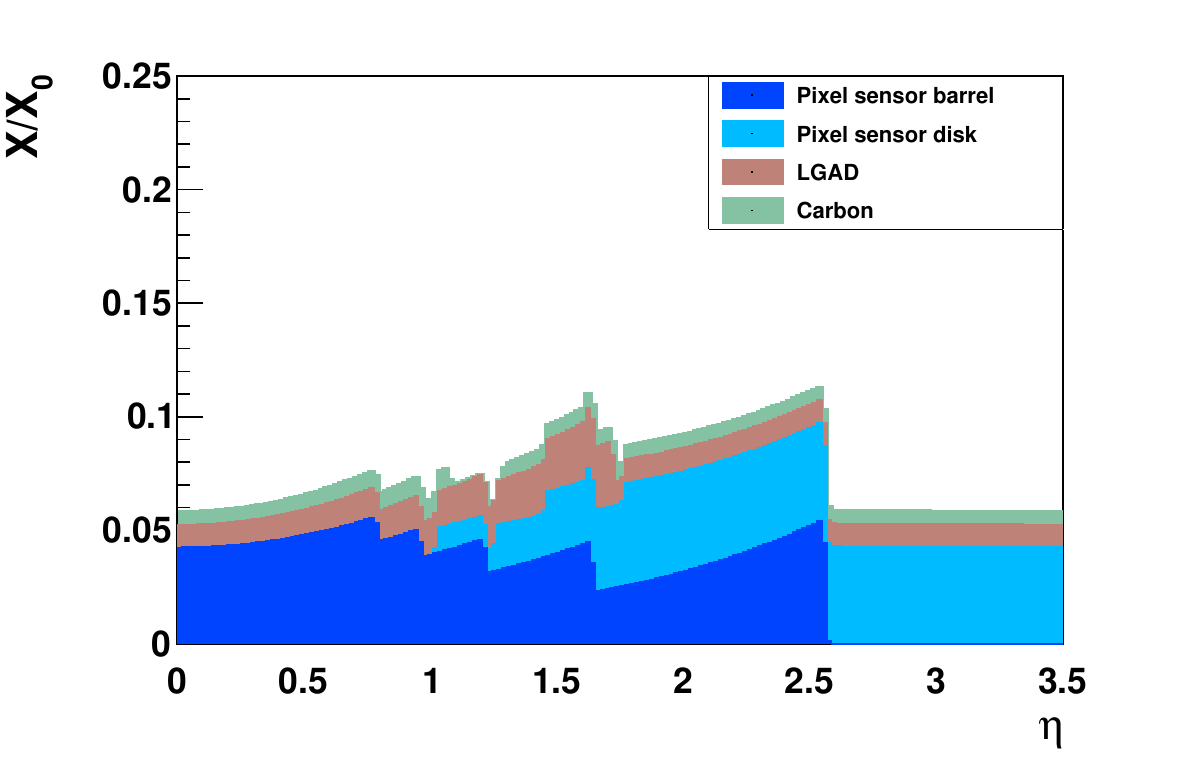}
\caption{\label{fig:mat_budget} {The material budget as a function of pseudorapidity for H-NS. The different colors show the contribution from the pixel sensor barrel (dark blue), pixel sensor disk (light blue), the LGAD detector (brown), and the carbon scattering layer (light green). 
}}
\end{figure}

\begin{figure}
\includegraphics[width=0.5\textwidth]{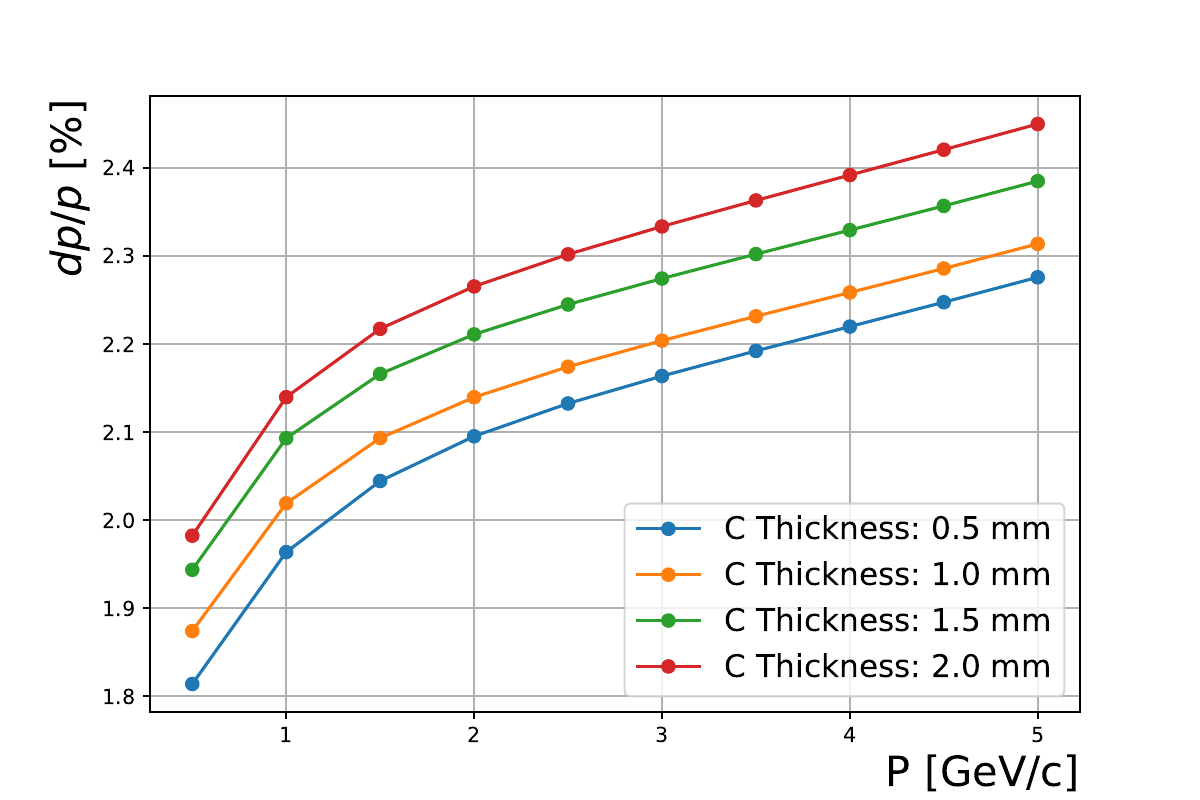}
\caption{\label{fig:mom_res_thickness} {Momentum resolution at different carbon thickness.
}}
\end{figure}

Adding an extra scattering layer of carbon is essential to measure the spin polarization of final-state nucleons. Its impact on conventional detector performance, such as tracking efficiency and momentum resolution, is investigated. A carbon layer with a thickness of 1 mm corresponds to a material budget below 1\% $\textrm{X/X}_0$. Figure~\ref{fig:mat_budget} shows the material budgets for each component in H-NS. The carbon layer contribution is shown in light green color. A tiny material budget from the new added carbon layer is seen. This carbon layer will introduce a very tiny influence on the tracking efficiency and momentum resolutions. As shown in Fig.~\ref{fig:mom_res_thickness}, the momentum resolution of charged particles is compared at different carbon thicknesses. A carbon layer of 1 mm thickness may deteriorate the momentum resolution by about 0.1\%. 

\section{Event selection}

% \begin{table}[h!]
% \begin{tabular}{|c|c|c|c|}
% \hline
%       & 1~GeV & 3~GeV & 5~GeV \\
% \hline
% \rowcolor{green} 85~$^{\circ}$ & 1~GeV & 2~GeV & 2~GeV \\
% \hline
% \rowcolor{green} 73~$^{\circ}$ & 1~GeV & 2~GeV & 2~GeV \\
% \hline
% \rowcolor{green} 61~$^{\circ}$ & 1~GeV & 2~GeV & 2~GeV \\
% \hline
% 20~$^{\circ}$ & 1~GeV & 2~GeV & 2~GeV \\
% \hline
% 10~$^{\circ}$ & 1~GeV & 2~GeV & 2~GeV \\
% \hline
% \end{tabular}
% %\caption{Summary of systematic uncertainties in the proton polarization measurement.}
% \label{table:theta_p}
% \end{table}

\begin{figure}
\includegraphics[width=0.5\textwidth]{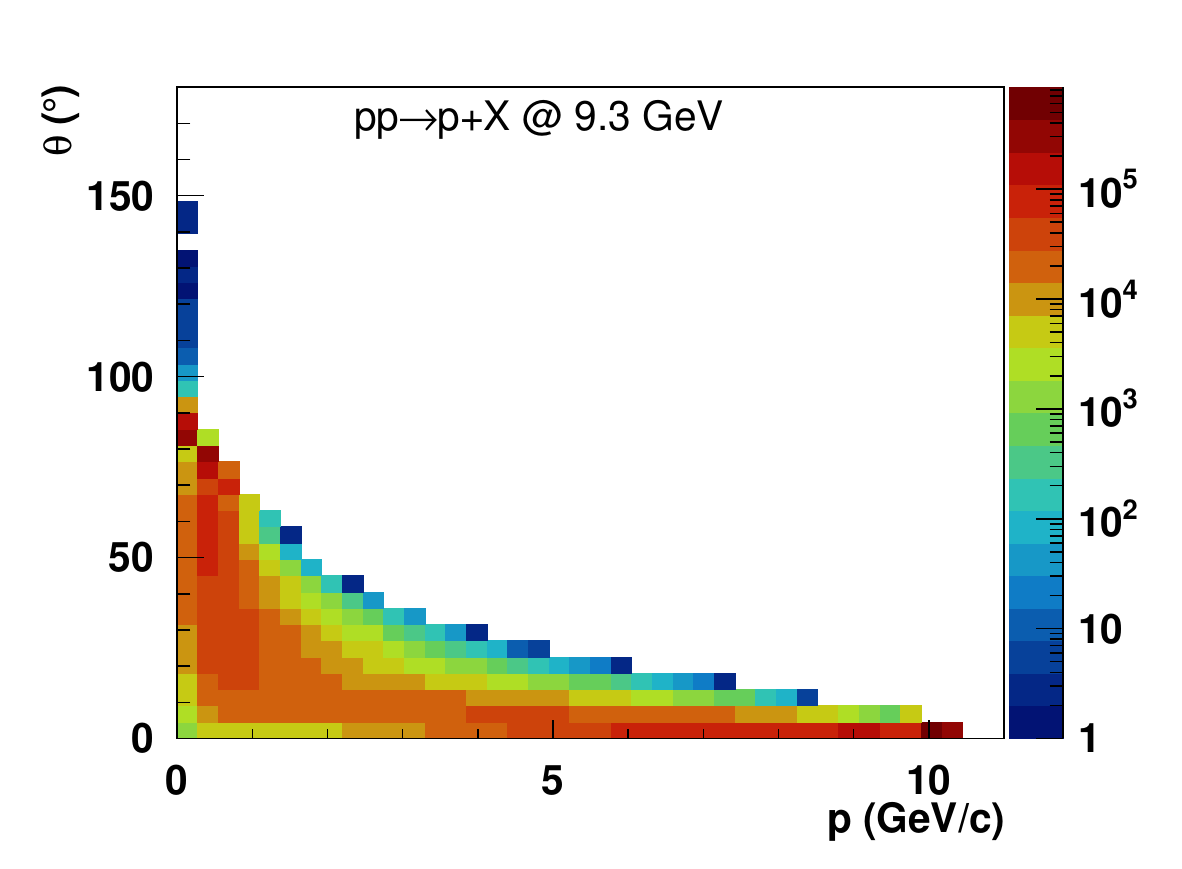}
\caption{\label{fig:kinematic_coverage_proton} {Momentum versus the polar angle of final-state proton produced in $pp$ collision with beam energy of 9.3 GeV.
}}
\end{figure}

\begin{figure}
\includegraphics[width=0.5\textwidth]{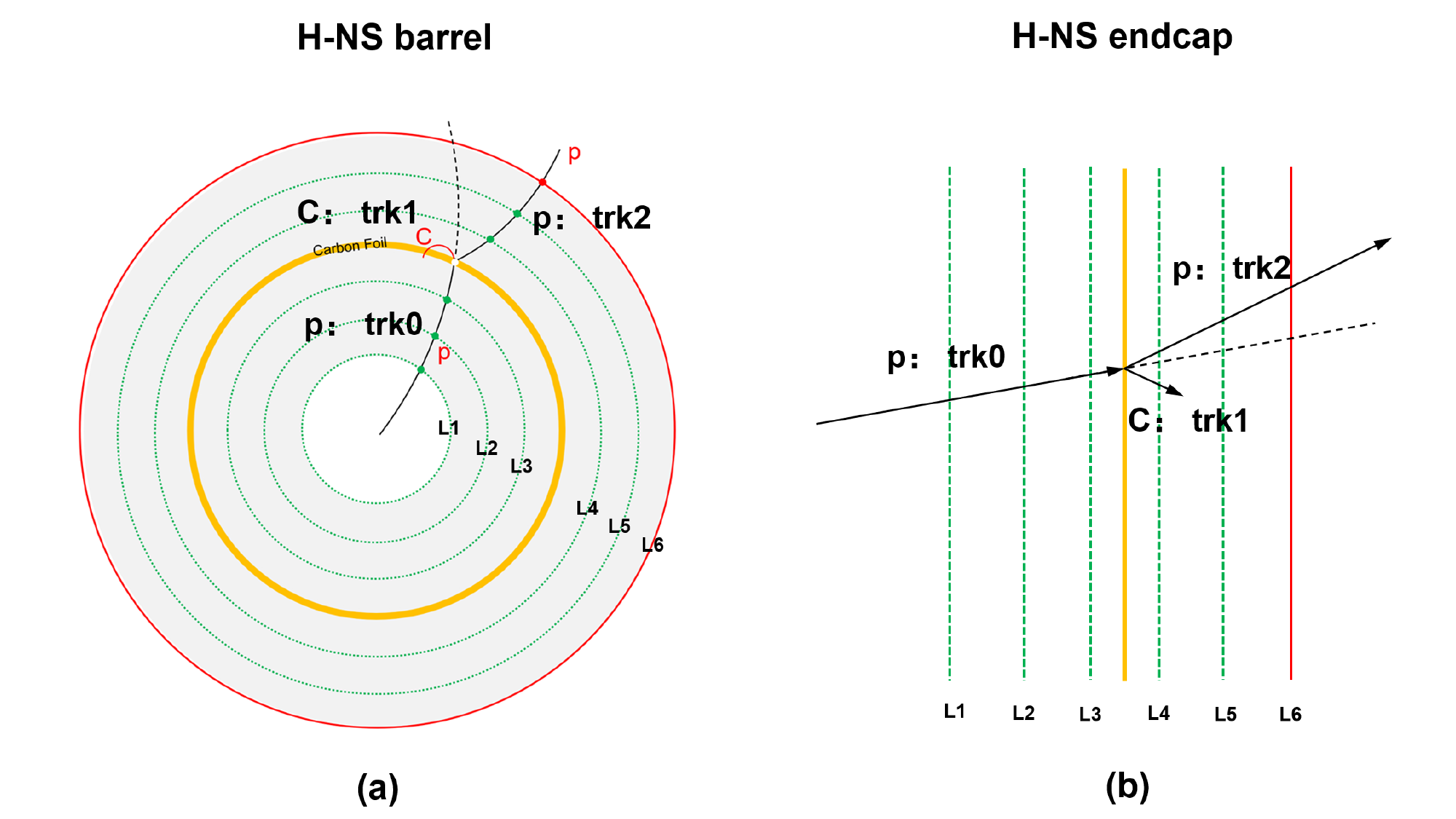}
\caption{\label{fig:pC_event_topology} {Illustration of proton elastic scattering on carbon layers in the barrel (a) and endcap (b) of the H-NS. Unscattered protons, contributing as the major background, are indicated with dashed curves.
}}
\end{figure}

For H-NS, one physics motivation is the systematic measurement of final-state proton polarization. Figure~\ref{fig:kinematic_coverage_proton} shows the momentum versus the polar angle of the final-state proton produced in $pp$ collision with a beam energy of 9.3 GeV.
In order to evaluate and optimize the performance of the proton polarimeter at the H-NS spectrometer, a full-process (realistic) simulation, including event generation, {\sc Geant4}~\cite{AGOSTINELLI2003250}  simulation, particle identification, signal selection, and polarization extraction, is performed. 
The sketches in Fig.~\ref{fig:pC_event_topology} are the main components of H-NS that are used to measure proton polarization.  The barrel (Fig.\ref{fig:pC_event_topology}a) and endcap (Fig.\ref{fig:pC_event_topology}b) parts are similar; both are composed of a carbon layer as the polarimeter target, sandwiched between two groups of tracking arrays, each of which consists of three layers of silicon pixel detectors. 
The proton-carbon elastic scattering takes place at the carbon layers. The inner solid line represents the path of the incident proton (trk0), while the scattered proton (trk2) and the recoil carbon nucleus (trk1) are shown with the solid curves on the outer side of the carbon layer. 
The recoil carbon nuclei have such small kinematic energies that they even cannot reach the nearby silicon layer. In other words, they are invisible to our detector.

Combining the information provided by the tracking system and the TOF detector (or possibly having a Cherenkov detector), protons can be distinguished from other particles like pions and kaons in the H-NS kinematic coverage. The performance of the proton identification is being optimized within the framework of the particle identification (PID) system and will not be explained in this paper.
%Combining the information provided by the tracking system and the TOF detector, one can distinguish protons from other particles like pions and kaons with a high efficiency (\textcolor{red}{XXX}) and a low background-to-signal ratio (\textcolor{red}{XXX}).
One dominant background comes from the protons that pass through the target layer without scattering. 
Illustrated by the dashed curves in Fig.~\ref{fig:pC_event_topology}, these unscattered protons can mimic elastic scattering events of low momentum transfer. To evaluate and optimize the polarimeter function, a series of MC simulation are performed at H-NS. Table~\ref{table:theta_p} lists the proton polar angle and momentum points in the simulation. For each ($p$, $\theta$) bin, 0.2 million $p\textrm{C}$ elastic events (signal) and 10 million unscattered events (background) are generated.
%All protons from signal events take the nominal momentum (p) and angle ($\theta$), while the background protons are sampled in the vicinity \textcolor{red}{(specific range?)} of the nominal values. 

\begin{table}[h!]
\begin{tabular}{|c|c|c|c|}
\hline
% \diagbox{angle}{}{energy} & \multicolumn{1}{l|}{1~GeV} & \multicolumn{1}{l|}{3~GeV} & 5~GeV \\ 
\diagbox{angle}{region}{energy} & 1~GeV/{\it c} & 3~GeV/{\it c} & 5~GeV/{\it c} \\ 
\hline
85$^{\circ}$ & \multicolumn{3}{c|}{\multirow{3}{*}{barrel}}           \\ \cline{1-1}
73$^{\circ}$ & \multicolumn{3}{c|}{}                            \\ \cline{1-1}
61$^{\circ}$ & \multicolumn{3}{c|}{}                            \\ 
 \hline
20$^{\circ}$ & \multicolumn{3}{c|}{\multirow{2}{*}{endcap}}           \\ \cline{1-1}
10$^{\circ}$ & \multicolumn{3}{c|}{}                            \\ 
 \hline
\end{tabular}
\caption{Proton polar angle-momentum points where MC simulations are performed for the study of proton final-state polarimetry at H-NS. Large and small scattering angles correspond to the barrel and endcap parts of the spectrometer, respectively. }
\label{table:theta_p}
\end{table}

After simulating particle propagation and signal digitization with the H-NS software, a valid hit is required at each silicon layer (L1, ..., L6). The tracks of charged particles are reconstructed with the GenFit package~\cite{HOPPNER2010518}, which essentially fits a cylindrical spiral to the hit positions, taking into account magnetic bending and multiple scattering.  
The fitting $\chi^2$ sets the criterion for judging whether a valid track can be reconstructed from a group of hits or not. 
Our aim is to find out a set of selection cuts to exclude unscattered proton events.

\begin{figure}
\includegraphics[width=0.5\textwidth]{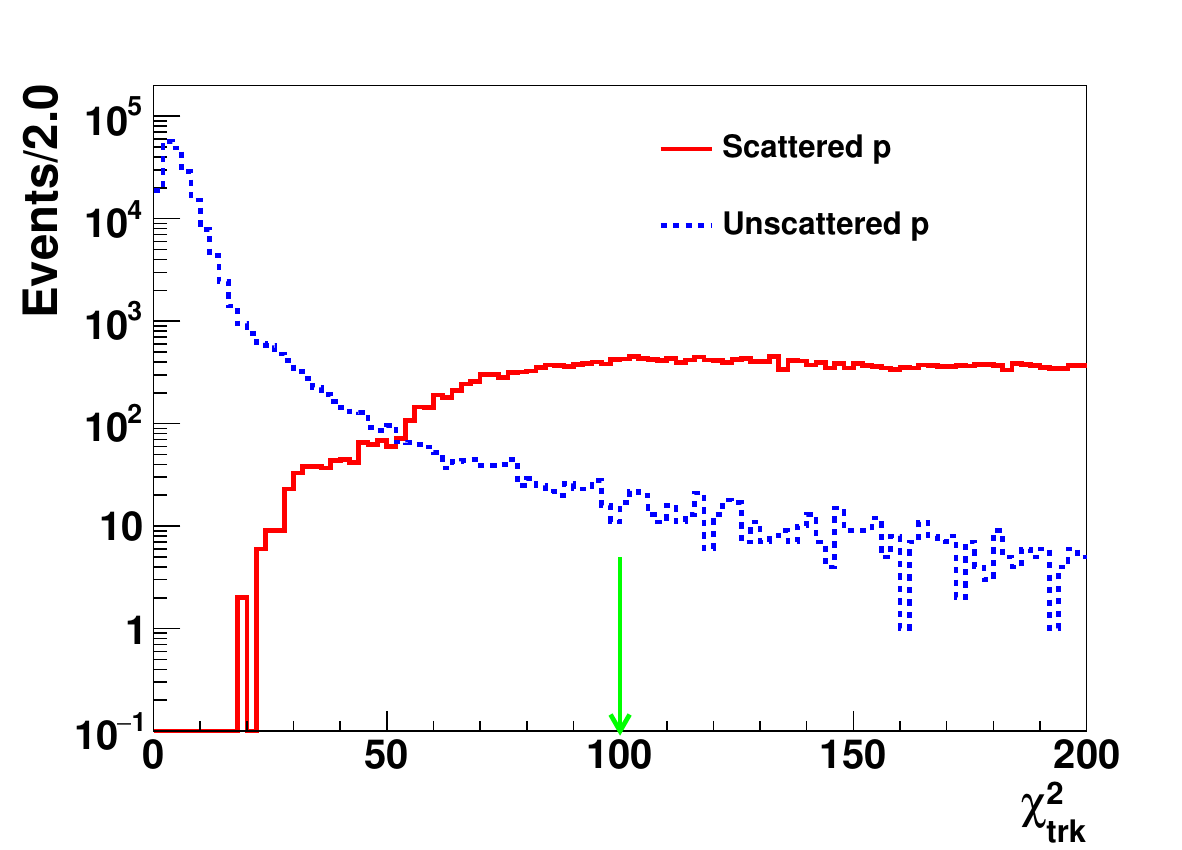}
\caption{\label{fig:chi2_cut} {$\chi^2$ of full track. The blue dashed line represents the unscattered proton with a small $\chi^2_{\rm trk}$ as anticipated. The red solid line is the scattered proton with a large $\chi^2_{\rm trk}$. The green arrow indicates the cut where $\chi^2_{\rm trk} \textgreater 100$.
}}
\end{figure}

We first assume that all six hits are produced by a single particle without scattering and perform a GenFit tracking procedure. The resultant $\chi^2_{\rm trk}$ is plotted in Fig.~\ref{fig:chi2_cut}. Applying a cut $\chi^2_{\rm trk} \textgreater 100$, 99.35\% of the unscattered proton events can be rejected.
Next, the hits on the inner layers (L1, L2, L3) and the hits on the outer layers (L4, L5, L6) are assigned to two individual particles, i.e. the incident and the scattered protons.
Consequently, two independent tracking procedures are performed separately.

\begin{figure}
\includegraphics[width=0.5\textwidth]{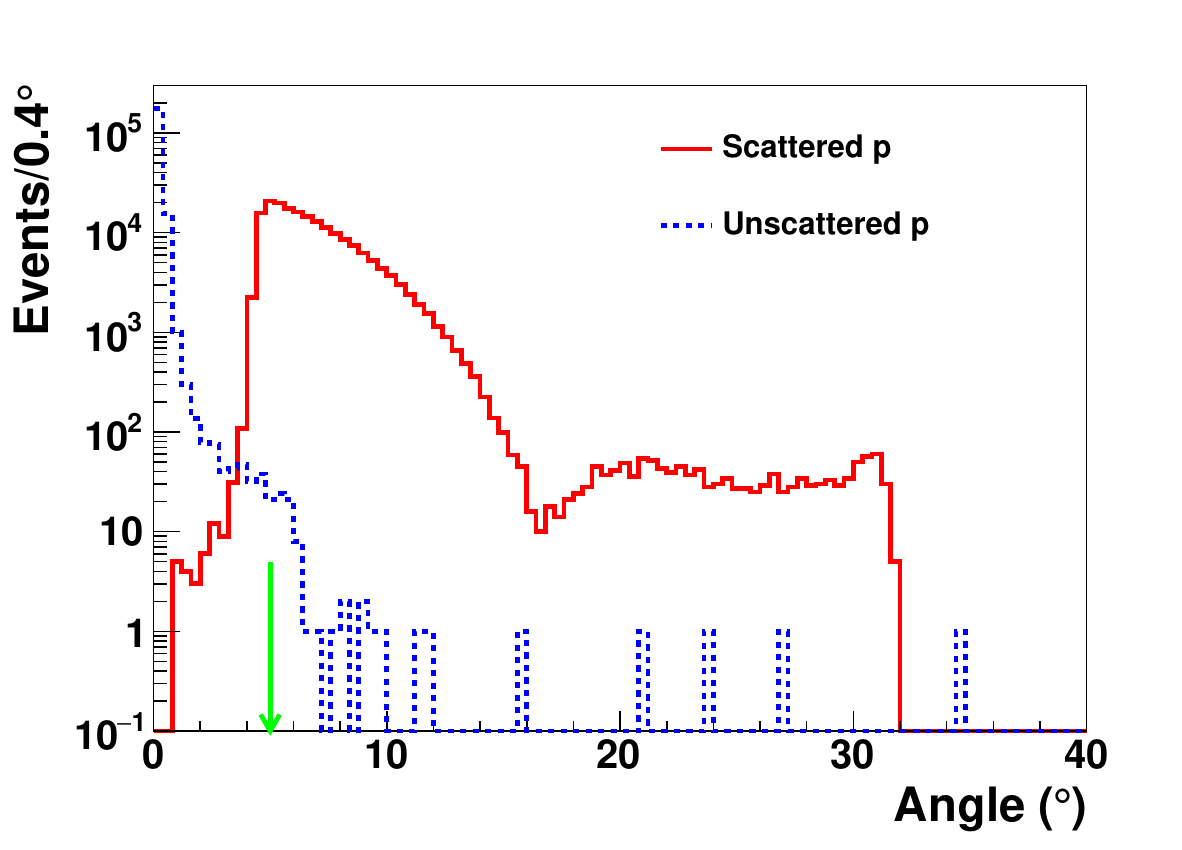}
\caption{\label{fig:angle_cut} {Angle between two track segments. The blue dashed line represents the unscattered proton with a small angle as anticipated. The red solid line is the scattered proton with a large angle. The green arrow shows the cut where $\theta_{\rm sc}\textgreater 5^{\circ}$.
}}
\end{figure}

Figure~\ref{fig:angle_cut} shows the proton scattering angle after the reconstruction of the track, defined as the tangent lines of the inner and outer tracks at their crossing point.
It is clear that separate tracks reconstructed from hits produced by a single unscattered particle tend to align with each other. Therefore, a single cut on proton scattering angle ($\theta_{\rm sc} \textgreater 5^{\circ}$) will exclude 99.96\% of unscattered protons.
Another selection cut is imposed on the tracking crossing point. Proton-carbon scattering only happens on the carbon target; therefore, by taking into account tracking resolution, the inner and outer tracks are expected to cross in a close vicinity of the carbon layer.   
\begin{figure}
\includegraphics[width=0.5\textwidth]{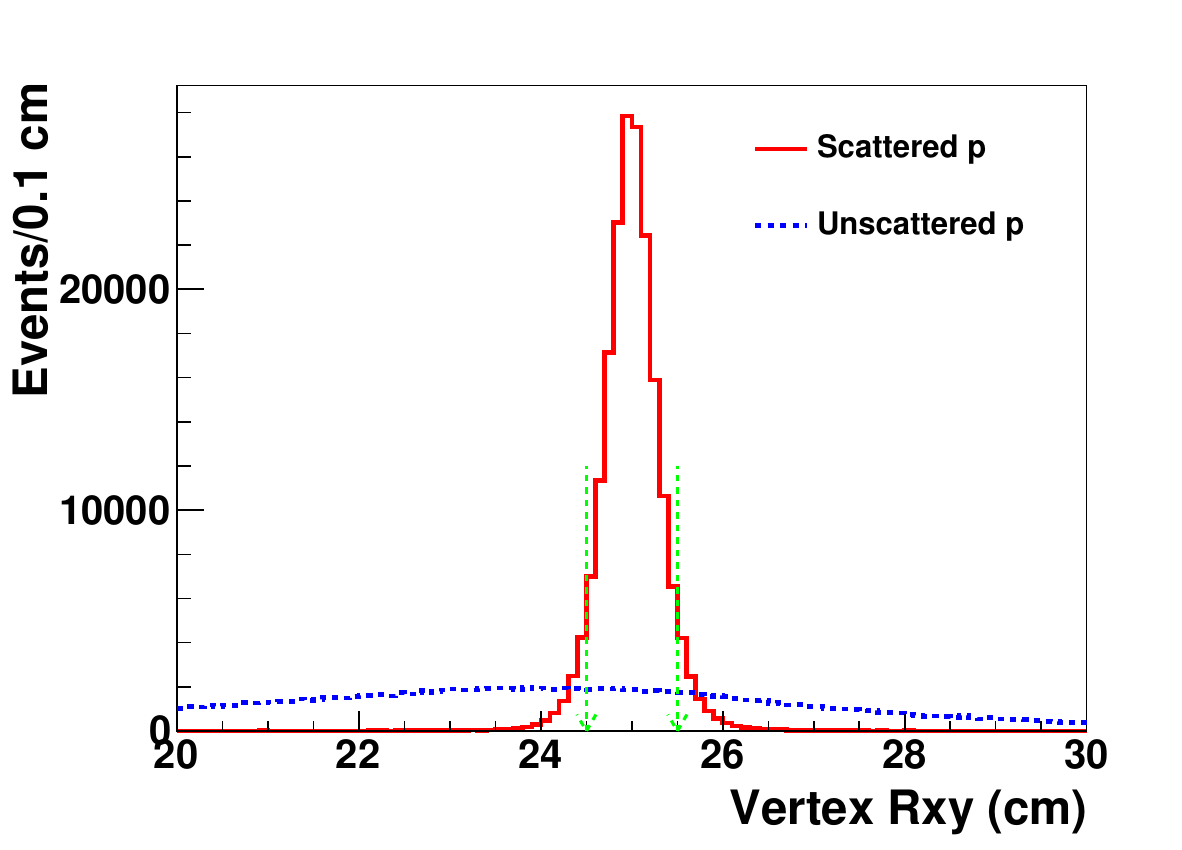}
\caption{\label{fig:evtsel_Rxy} {The radial position ($R_{xy}$) of the track cross for $p\textrm{C}$ scattering events (solid line in red) and unscattered tracks (dashed line in blue) at barrel detector. The $p\textrm{C}$ scattering signal region is selected between the two green arrows.
}}
\end{figure}
The track cross point is reconstructed with the RAVE package~\cite{WALTENBERGER2007549}. As shown in Fig.~\ref{fig:evtsel_Rxy}, the radial position ($R_{xy}$) of the track cross for the $p\textrm{C}$ scattering events at the barrel detector follows a narrow Gaussian distribution ($\sigma = 0.29~{\rm cm}$), while the unscattered protons are spread much more widely.  
In general, by applying all the cuts introduced above together, a selection efficiency of 70\% and a background suppression ratio of $4\times10^{-6}$ are achieved. Given the lower cross section of the scattered signal, the background-to-signal ratio below 1\% is expected. Another background source is inelastic scattering, where the carbon nucleus is broken. Most of these inelastic backgrounds can be rejected by the multiplicity of tracks. From a {\sc Geant4} simulation, the inelastic scattering background may survive and contaminate the signal at a few percentage levels. The selected signal events will be of high purity, which is important in extracting the polarization eventually.

% \begin{figure}
% \includegraphics[width=0.45\textwidth]{mom_resolution.pdf}
% \caption{\label{fig:mom_res} {Momentum resolution using 3 outer layers after scattering. \sout{\textcolor{blue}{To CX: please combine the difference distributions of p, $\theta$ and $\phi$.}
% }}}
% \end{figure}
%

%Comparison between the generated and reconstructed tracks is made for the recoil proton to estimate the momentum resolution ($\sigma_p$), polar angle resolution ($\sigma_\theta$) and azimuthal angle resolution ($\sigma_\phi$), which are obtained as the distribution widths of $p^{rec}-p^{gen}$, $\theta^{rec}-\theta^{gen}$ and $\phi^{rec}-\phi^{gen}$ (see Figure~\ref{fig:mom_res}).

\begin{figure}
\includegraphics[width=0.5\textwidth]{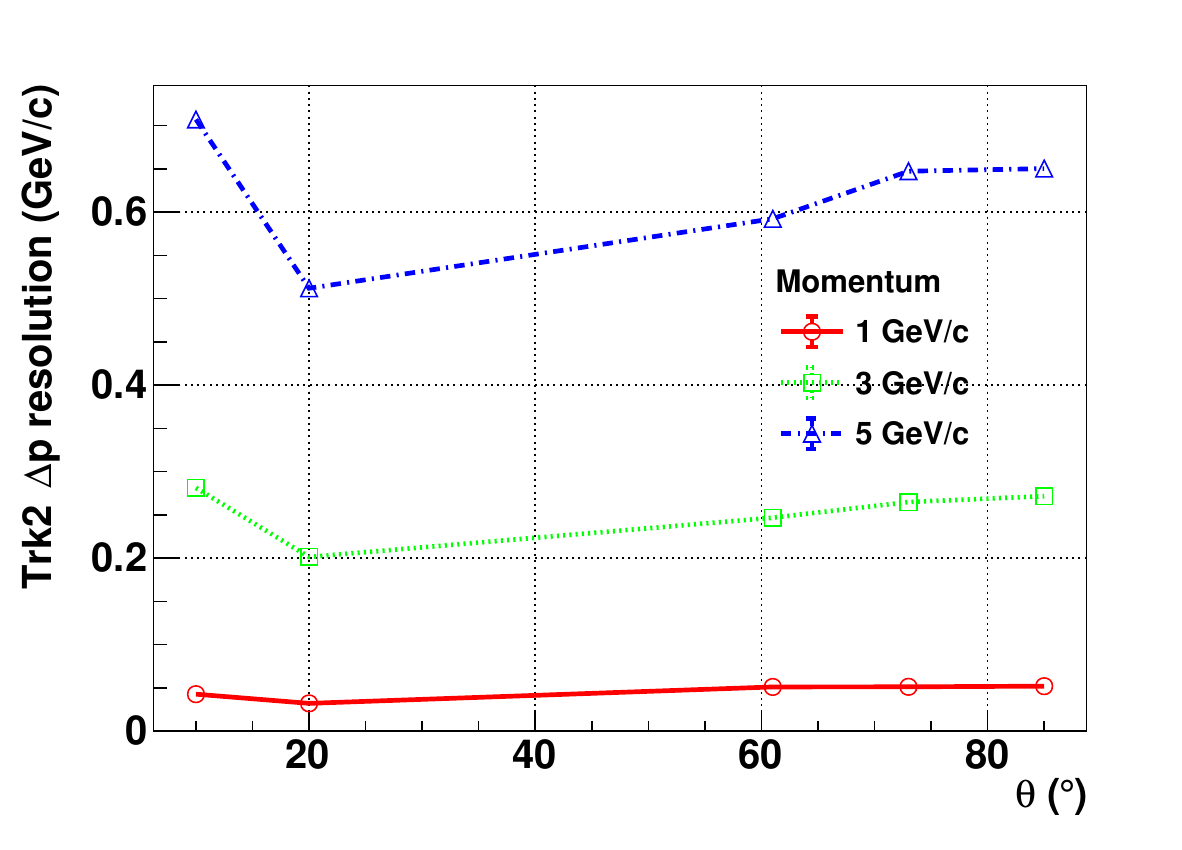}
\caption{\label{fig:res_mom_recoil_p} {Momentum resolution of the scattered proton reconstructed with the outer three tracking layers for incident proton with momentum of 1 GeV/{\it c}, 3 GeV/{\it c} and 5 GeV/{\it c}, respectively.
}}
\end{figure}

\begin{figure}
\includegraphics[width=0.5\textwidth]{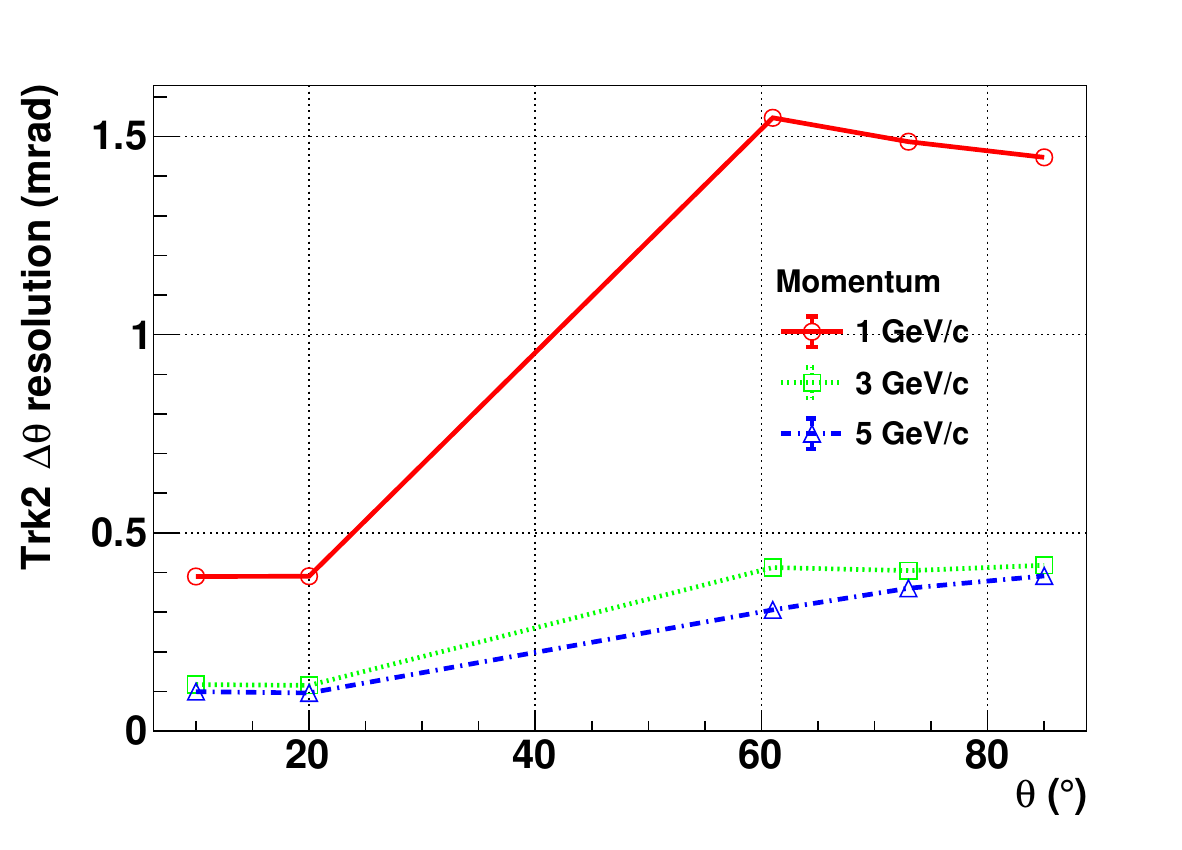}
\caption{\label{fig:res_theta_recoil_p} {Resolution on $\Delta\theta$ of the scattered proton reconstructed with the outer three tracking layers for incident proton with momentum of 1 GeV/{\it c}, 3 GeV/{\it c} and 5 GeV/{\it c}, respectively.
}}
\end{figure}

\begin{figure}
\includegraphics[width=0.5\textwidth]{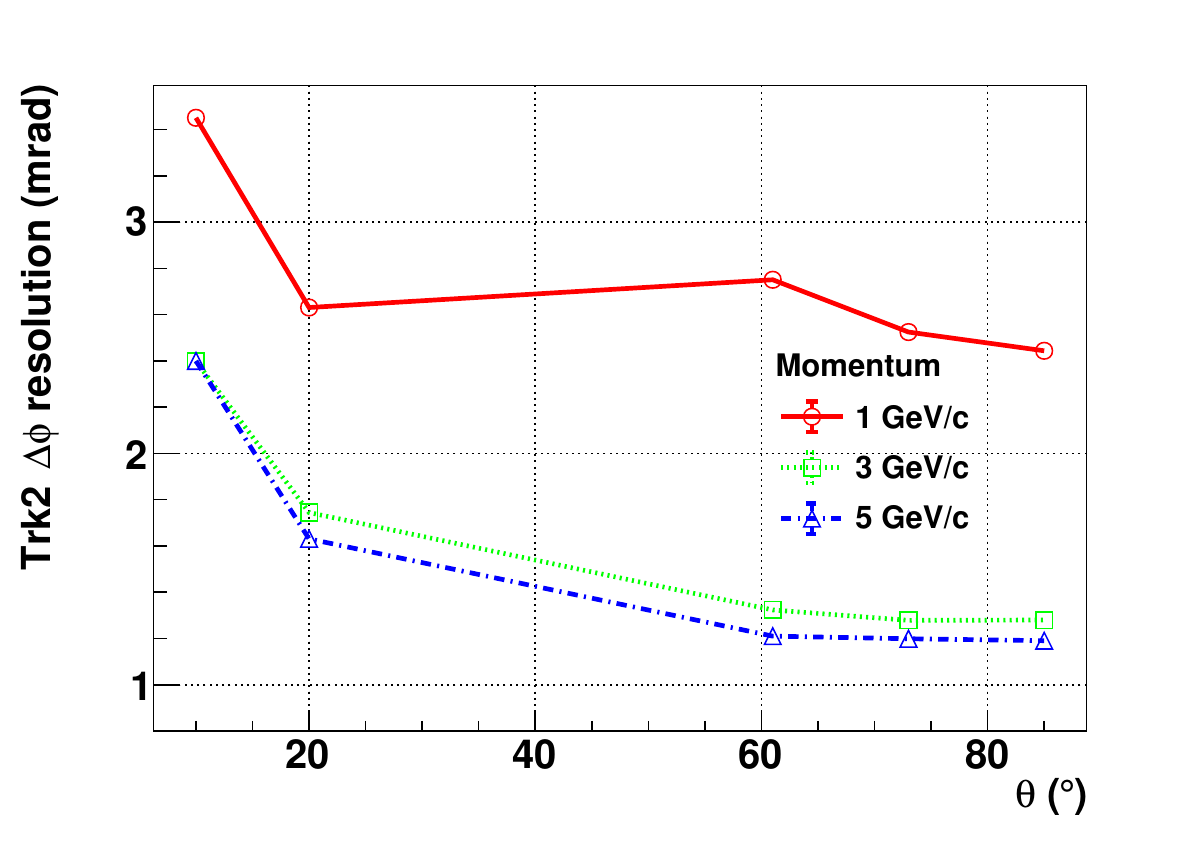}
\caption{\label{fig:res_phi_recoil_p} {Resolution on $\Delta\phi$ of the scattered proton reconstructed with the outer three tracking layers for incident proton with momentum of 1 GeV/{\it c}, 3 GeV/{\it c} and 5 GeV/{\it c}, respectively.
}}
\end{figure}

The MC simulation discussed above is carried out for every polar angle and momentum combination listed in Table~\ref{table:theta_p}, and as a result, the event selection efficiency, background suppression ratio, and resolution of the scattered proton momentum are obtained for each combination. The momentum resolution $\Delta p$ (difference between the reconstructed proton momentum and the generated proton momentum) of the scattered proton reconstructed with the outer three tracking layers is shown in Fig.~\ref{fig:res_mom_recoil_p}. The momentum resolution gets worse with higher momentum, as expected. The resolution on $\Delta\theta$ (difference between the reconstructed polar angle and the generated polar angle) of the scattered proton is shown in Fig.~\ref{fig:res_theta_recoil_p}, and the resolution on $\Delta\phi$ (difference between the reconstructed azimuthal angle and the generated azimuthal angle) of the scattered proton is shown in Fig.~\ref{fig:res_phi_recoil_p}. Here, a very good angular resolution on $\Delta\theta$ and $\Delta\phi$, at a level of a few milliradians, is seen. These accurate angular measurements are essential for the extraction of polarization.

% \begin{figure}
% \includegraphics[width=0.45\textwidth]{pC_Xsc_Ay_500MeV.png}
% \caption{\label{fig:pC_Xsc_Ay} {The differential cross-section $\frac{d\sigma}{d\Omega}$ (upper) and analyzing power $A_y$ (lower) of $\vec{p}C$ elastic scattering as functions of the scattered proton angle in the c.m. system, at the incident proton energy of 500~MeV. Plot adopted from Ref~\cite{PhysRevC.41.1651}.
% }}
% \end{figure}

% \subsection{Event generation}

% \subsection{Signal selection}

% \subsection{Polarization extraction}

% \begin{figure}
% \includegraphics[width=0.45\textwidth]{phi_resolution.pdf}
% \caption{\label{fig:phi_res} {$\phi$ resolution using 3 outer layers after scattering..
% }}
% \end{figure}

% \begin{figure}
% \includegraphics[width=0.45\textwidth]{vertex_rec.pdf}
% \caption{\label{fig:Hvertex_rec} {Reconstructed vertex by two track segments.
% }}
% \end{figure}

\section{Probability of $p\textrm{C}$ elastic scattering}
\begin{figure}
\includegraphics[width=0.47\textwidth]{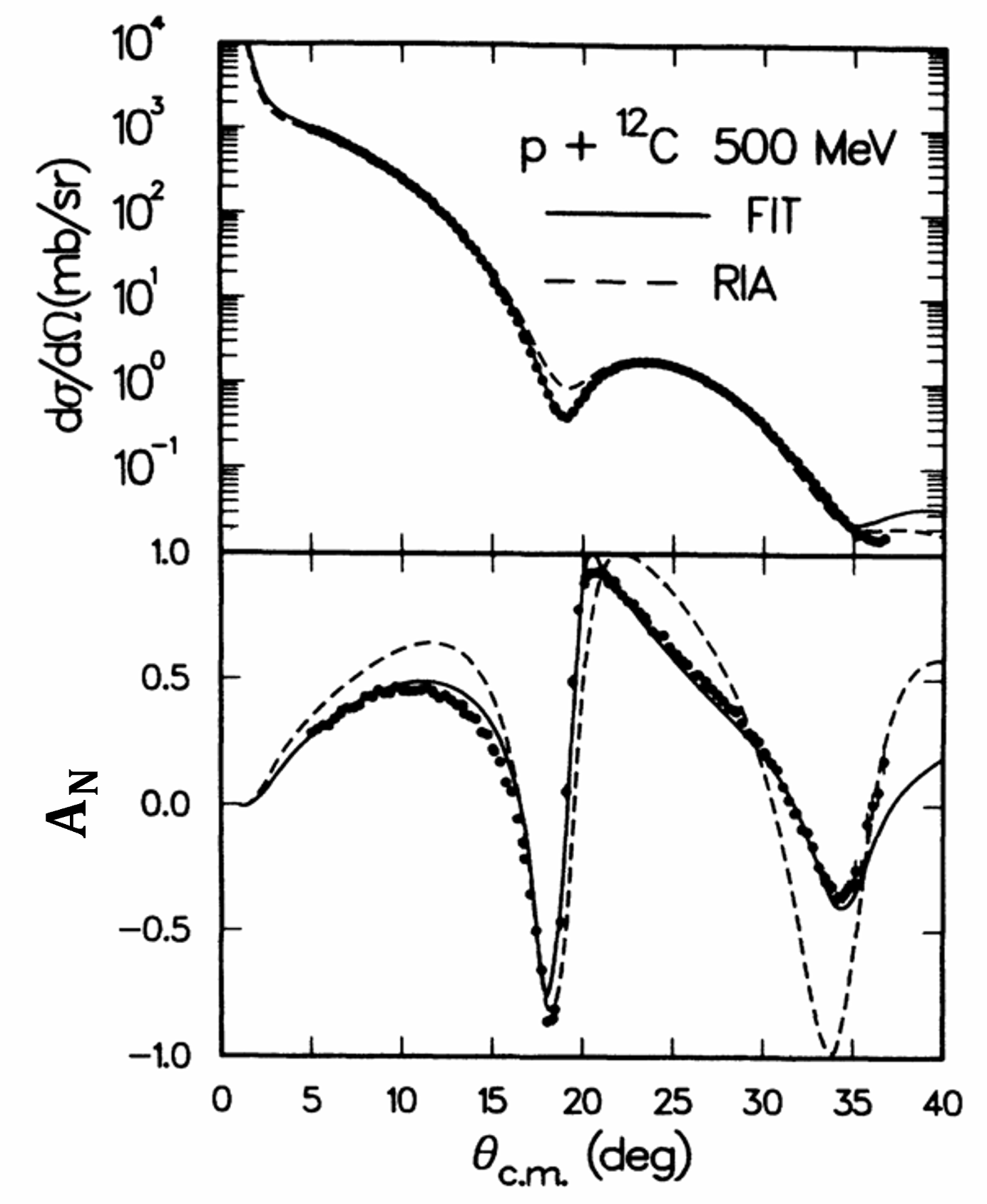}
\caption{\label{fig:pC_Xsc_Ay} {The differential cross-section $\frac{d\sigma}{d\Omega}$ (upper) and analyzing power $A_y$ (lower) of $p\textrm{C}$ elastic scattering as functions of the scattered proton angle in the c.m. system, at the incident proton energy of 500~MeV. Plot adapted from Ref~\cite{PhysRevC.41.1651}.
}}
\end{figure}
%As shown in Ref~\cite{10.1093/ptep/ptx116}, the pC reaction cross section decreases with energy from 10~MeV to 100~MeV and remains almost constant up to 1~GeV. 
As shown in Ref~\cite{10.1093/ptep/ptx116}, the $p\textrm{C}$ reaction cross section is almost constant between 100~MeV and 1~GeV. 
Therefore, it is reasonable to make a conservative estimate of the probability of $p\textrm{C}$ scattering with the cross section measured at a kinetic energy of 500~MeV~\cite{PhysRevC.41.1651}.
MC simulation reveals that $p\textrm{C}$ elastic scattering events are well separated from the background when the proton scattering angle $\theta_{\rm sc}$ is greater than 5 degrees and the event ratio becomes negligibly small above 15 degrees (see Figure~\ref{fig:angle_cut}). 
At 500 MeV, the integral cross section $\sigma_{Int}$ between 5 and 20 degrees is about 55~mb. 
The areal density of carbon nucleus in a target made of a pure carbon layer with a thickness of $\mathcal{D}$~cm is $\Sigma_C = \frac{\mathcal{D} \cdot \rho_c}{M_c } \cdot N_A = \mathcal{D} \cdot 7.9 \times 10^{22}~\rm{cm^{-2}}$, 
 where $\rho_c=1.57~\rm{g/cm^3}$ is the density of pure carbon, $M_c = 12~\rm{g/mol}$ the Molar mass of $\rm{^{12}C}$, and $N_A$ the Avogadro number.
If the target thickness is $\mathcal{D}=1~\rm{mm}$, the probability of $p\textrm{C}$ elastic scattering is $\mathcal{P} = \sigma_{Int} \cdot \Sigma_c = \mathcal{D} \cdot 43.45 \times 10^{-4} \approx 4.3 \times 10^{-4}$. This tiny scattering probability is the principle of the technique, which measures the polarization with a tiny fraction of nucleons. High-luminosity machines like H-NS will produce vast quantities of nucleons. Consequently, the low scattering probability inherent to this technique poses no issue for statistical precision.

%This principle also explains the negligible impact on the standard performance of the spectrometer, as we have explained in Fig.~\ref{fig:mat_budget} and ~\ref{fig:mom_res_thickness}.

%
\section{Polarization extraction}
% Polarimeter performance is evaluated as follows. 
% \textcolor{red}{Sufficient} pC elastic scattering events are generated in the range of scattering angle covered by the outer tracking detectors of the H-NS spectrometer. 
% The polar angle distribution and the azimuthal angle distribution follow the differential cross section and the analyzing power (\textcolor{red}{Figure~\ref{fig:pC_Xsc_Ay}}), respectively. 
% The effective analyzing power $A_N^{eff}$, used to describe the azimuthal distribution, is taken as the average of $A_{N}(\theta)$, weighted by the differential cross section $\frac{d\sigma}{d\Omega}$ within the polar angle uncertainty $\sigma_{\theta}$.
% The uncertainty in measuring the azimuthal angle is taken into account by convoluting the cosine azimuthal distribution with a Gaussian function with a width of $\sigma_{\phi}$.
% %
% \begin{equation}
% N(\phi) =  N_0\left[1+e^{-\frac{\sigma_{\phi}^2}{2}}\mathcal{P}_y A_N^{eff}\cos\phi\right]
% \label{eq:conv_pol_XSC}
% \end{equation}
% %
% The proton polarization $\mathcal{P}_y$ is assumed to be 100\%. 
% By fitting the generated azimuthal distribution with Equation~\ref{eq:pol_XSC}, both the polarization and its uncertainty are obtained, as plotted in Figure-XXX as functions of the scattering angle for different proton energies.

In this section, we illustrate the procedure for extracting the final-state proton polarization and identify the key factors that would affect the experimental precision.  
Suppose that we collect $\mathcal{N}_{\rm tot}$ events of elastic scattering of $p\textrm{C}$ induced by protons emitted from the primary vertex at some certain angle and kinetic energy. Take proton with 500~MeV kinetic energy as example, which corresponds to a momentum of 1.06 GeV/c, the angular dependent cross section and analyzing power of $p\textrm{C}$ scattering have been measured~\cite{PhysRevC.41.1651} and shown in Fig.~\ref{fig:pC_Xsc_Ay}. 

\begin{figure}
\includegraphics[width=0.5\textwidth]{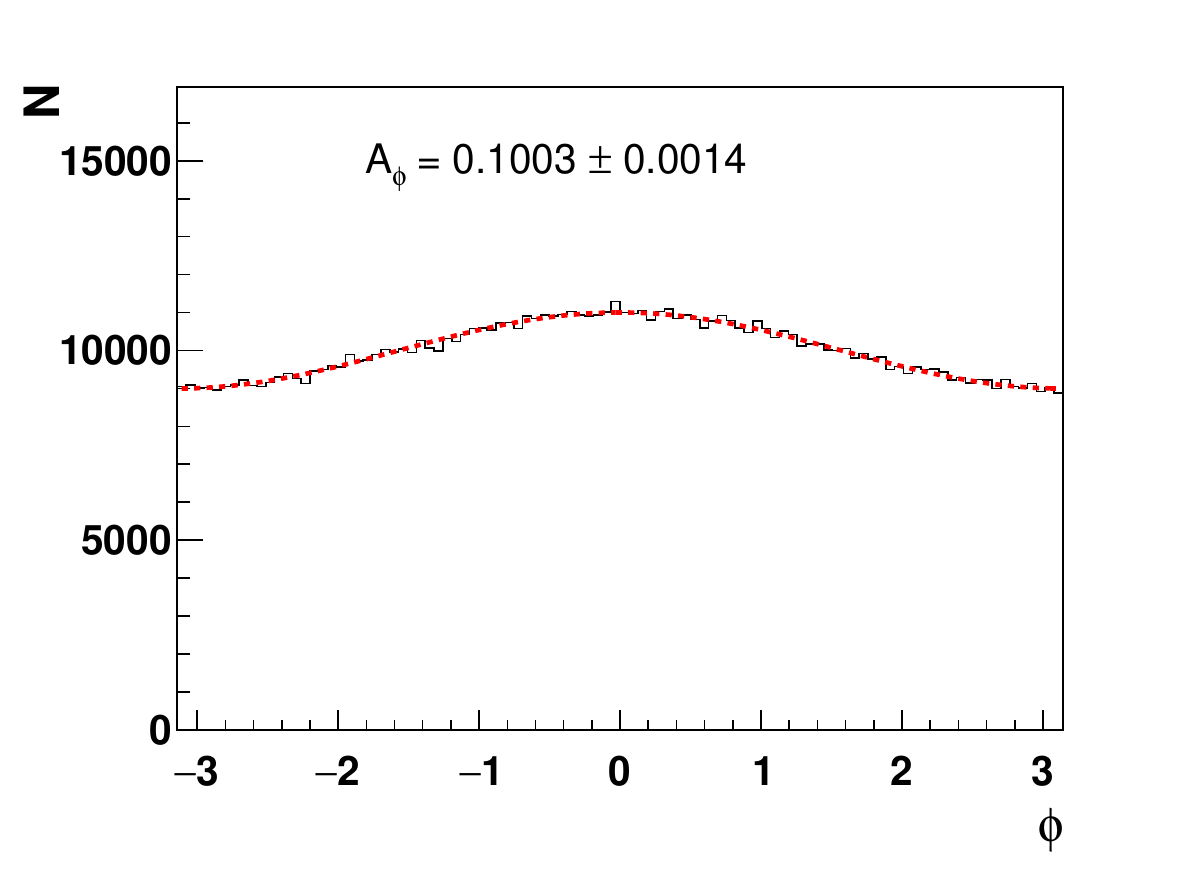}
\caption{\label{fig:Ass_IO} {With 1 million $p\textrm{C}$ elastic scattering events at H-NS, the extracted asymmetry is consistent with the input asymmetry of 0.1. 
}}
\end{figure}

The azimuthal distribution of the scattered protons in the scattering frame is described by
\begin{equation}
N(\phi) =  \frac{\mathcal{N}_{\rm tot}}{2\pi}\left[1 + \mathcal{P}_y A_N^{\rm ave}\cos\phi\right].
\label{eq:conv_pol_XSC}
\end{equation}
The azimuthal asymmetry, defined as 
\begin{equation}
A_{\phi} = \mathcal{P}_y \cdot A_N^{\rm ave},
\label{eq:Asy_phi}
\end{equation}
characterizes how much the polarized distribution deviates from the uniform unpolarized distribution. 
The average analyzing power $A_N^{\rm ave}$ is the average of $A_{N}(\theta)$, weighted by the differential cross section $\frac{d\sigma}{d\Omega}(\theta)$ in the angular range covered by the detector.
%The term $e^{-\frac{\sigma_{\phi}^2}{2}}$ accounts for the asymmetry suppression associated with the uncertainty in measuring the azimuthal angle.
$A_{\phi}$ is extracted by fitting the experimental azimuthal distribution, its statistical uncertainty is related to the total number of events as $\Delta A_{\phi} = \sqrt{\frac{2}{\mathcal{N}_{\rm tot}}}$. Figure~\ref{fig:Ass_IO} shows the reconstructed $\phi$ distribution for the input azimuthal asymmetry of 0.1. The extracted asymmetry is consistent with the input value.

The proton polarization $\mathcal{P}_y$ is then obtained by normalizing $A_{\phi}$ with the average analyzing power, and its relative uncertainty is expressed as
% %
% %
% \begin{equation}
% \frac{\Delta A_{\phi}}{A_{\phi}} = \frac{\pi/2}{\mathcal{P}_y \cdot A_N^{eff} \cdot \sqrt{\mathcal{N}_{tot}}}
% \label{eq:conv_pol_XSC}
% \end{equation}
% %
% %
% \begin{equation}
% \frac{\Delta \mathcal{P}_y}{\mathcal{P}_y} = \sqrt{\Bigg(\frac{\Delta A_{\phi}}{A_{\phi}}\Bigg)^2 + \Bigg(\frac{\Delta A_N^{eff}}{A_N^{eff}}\Bigg)^2}
% \label{eq:conv_pol_XSC}
% \end{equation}
% %
%
\begin{equation}
\frac{\Delta \mathcal{P}_y}{\mathcal{P}_y} = \sqrt{\frac{2}{\mathcal{N}_{\rm tot}} \Bigg(\frac{1}{\mathcal{P}_y \cdot A_N^{\rm ave}}\Bigg)^2 + \Bigg(\frac{\Delta A_N^{\rm ave}}{A_N^{\rm ave}}\Bigg)^2}.
\label{eq:conv_pol_XSC}
\end{equation}

The uncertainty in $\mathcal{P}_y$ depends on the statistics and the systematic uncertainty associated with the average analyzing power. In the future H-NS experiment, the 1 MHz event rate ensures that statistical uncertainty is negligible. Consequently, the precision of the transverse polarization measurement will be limited primarily by the systematic uncertainty associated with the average analyzing power. This makes a dedicated measurement of the analyzing power over a broader energy range essential for the experiment. Currently, the analyzing power of $p\textrm{C}$ scattering is primarily known for proton kinetic energies below 1 GeV. For higher energies, a critical data gap exists. We propose a novel, self-calibrating solution using polarized protons from hyperon decays, as available in H-NS data. As noted in Ref.~\cite{c642-1lzb}, daughter protons from hyperon decays are naturally produced with a high degree of transverse polarization. This source provides a practical, in-situ polarized proton beam to calibrate the angular-dependent analyzing power at high energies—a measurement previously unattainable.

\section{Summary}

In this paper, a novel technique is adopted for the H-NS experiment to add a polarimeter function to a general-purpose spectrometer. A detailed optimization process successfully established a robust capability for nucleon polarization measurement, while preserving the full performance of conventional detector systems. This design enables the H-NS experiment to measure final-state proton polarization. Systematically measuring this polarization across different collision energies, transverse momenta, and longitudinal momentum fractions of the final-state proton will provide crucial insights into the mechanisms behind proton polarization and spin structure of the nucleon. Moreover, by comparing these results with $\Lambda$ hyperon polarization data, the experiment will clarify the specific role of the strange quark in hyperon polarization. This design has been demonstrated in a low-track-multiplicity, low-occupancy environment. For future applications in heavy-ion collisions with extremely high multiplicity, or in the higher luminosity situation with pile-up of events, the tracking algorithm must be optimized to adapt to these challenging conditions.

The integration of the nucleon polarimeter is highly flexible. Its scattering target is tunable both in thickness (to control the scattering rate) and in material. For instance, by replacing carbon with a hydrogen-rich material like polyethylene, the polarimeter can exploit $pp$ scattering. This is advantageous, as it leverages the well-established and extensive dataset on $pp$ analyzing power.  
The inaugural integration of a polarimeter into a general-purpose spectrometer sets a valuable precedent for future experiments, such as the EIC~\cite{AbdulKhalek:2021gbh}, CEPC, STCF~\cite{Achasov:2023gey} and EicC~\cite{Anderle:2021wcy}, etc. By combining spin‑polarization measurements with the four‑momentum of final‑state particles, this setup enables deeper insight into the fundamental physics of nuclear and particle interactions.

%CEPC~\cite{CEPC},
%\textcolor{red}{(Same as the Abstract) This first integration of a polarimeter into a general-purpose spectrometer provides a valuable benchmark for future experiments. By combining spin polarization data with the four-momentum of final-state particles, we can achieve a more comprehensive understanding of the physics in nuclear and particle physics research.}

% 

\begin{acknowledgments}
This work is supported in part by the National Key Research and Development Program of China under Contract No. 2023YFA1606800, 2024YFA1611000, the National Natural Science Foundation of China (NSFC) under Contract No. 11975278, the Natural Science Foundation of Beijing, China under Contract No. JQ22002.
%Y.T.L thanks Yi Yin, Changzheng Yuan, Nu Xu for helpful discussions. B.X.G thank Igor Strakovsky and Ronald Workman for providing up-to-date SAID solutions for the analyzing power of proton-proton scattering. 
 
\end{acknowledgments}

% Create the reference section using BibTeX:
\bibliography{polarimeter}

@article{YANG2013263,
title = "{High Intensity heavy ion Accelerator Facility (HIAF) in China}",
journal = {Nucl. Instr. Meth. B},
volume = {317},
pages = {263-265},
year = {2013},
issn = {0168-583X},
doi = {https://doi.org/10.1016/j.nimb.2013.08.046},
url = {https://www.sciencedirect.com/science/article/pii/S0168583X13009877},
author = {J.C. Yang and others}
}

@article{xucao2024,
  title = {Production and decay of hyperons in a transversely polarized electron-positron collider},
  author = {Cao, Xu and Liang, Yu-Tie and Ping, Rong-Gang},
  journal = {Phys. Rev. D},
  volume = {110},
  issue = {1},
  pages = {014035},
  numpages = {11},
  year = {2024},
  month = {Jul},
  publisher = {American Physical Society},
  doi = {10.1103/PhysRevD.110.014035},
  url = {https://link.aps.org/doi/10.1103/PhysRevD.110.014035}
}

@article{HOPPNER2010518,
title = "{A novel generic framework for track fitting in complex detector systems}",
journal = {Nucl. Instr. Meth. A},
volume = {620},
number = {2},
pages = {518-525},
year = {2010},
issn = {0168-9002},
doi = {https://doi.org/10.1016/j.nima.2010.03.136},
url = {https://www.sciencedirect.com/science/article/pii/S0168900210007473},
author = {C. Höppner and others}
}

@article{WALTENBERGER2007549,
title = "{RAVE—a Detector-independent vertex reconstruction toolkit}",
journal = {Nucl. Instr. Meth. A},
volume = {581},
number = {1},
pages = {549-552},
year = {2007},
note = {VCI 2007},
issn = {0168-9002},
doi = {https://doi.org/10.1016/j.nima.2007.08.048},
url = {https://www.sciencedirect.com/science/article/pii/S0168900207017196},
author = {W. Waltenberger and others}
}

@article{Sadrozinski:2017qpv,
    author = {Sadrozinski, Hartmut F. W. and others},
    title = {4\uppercase{D} tracking with ultra-fast silicon detectors},
    eprint = {1704.08666},
    archivePrefix = {arXiv},
    primaryClass = {physics.ins-det},
    doi = {https://doi.org/10.1088/1361-6633/aa94d3},
    url = {https://iopscience.iop.org/article/10.1088/1361-6633/aa94d3},
    journal = {Rept. Prog. Phys.},
    volume = {81},
    number = {2},
    pages = {026101},
    year = {2018}
}

@article{RTurchetta_2006,
doi = {10.1088/1748-0221/1/08/P08004},
url = {https://doi.org/10.1088/1748-0221/1/08/P08004},
year = {2006},
month = {aug},
publisher = {},
volume = {1},
number = {08},
pages = {P08004},
author = {R Turchetta},
title = {\uppercase{CMOS} \uppercase{M}onolithic \uppercase{A}ctive \uppercase{P}ixel \uppercase{S}ensors (\uppercase{MAPS}) for future vertex detectors},
journal = {JINST}
}

@article{BESIII:2018cnd,
    author = "Ablikim, M. and others",
    collaboration = "BESIII Collaboration",
    title = "{Polarization and Entanglement in Baryon-Antibaryon Pair Production in Electron-Positron Annihilation}",
    eprint = "1808.08917",
    archivePrefix = "arXiv",
    primaryClass = "hep-ex",
    doi = "10.1038/s41567-019-0494-8",
    journal = "Nature Phys.",
    volume = "15",
    pages = "631--634",
    year = "2019"
}

@article{Achasov:2023gey,
    author = "Achasov, M. and others",
    title = "{STCF conceptual design report (Volume 1): Physics \& detector}",
    eprint = "2303.15790",
    archivePrefix = "arXiv",
    primaryClass = "hep-ex",
    doi = "10.1007/s11467-023-1333-z",
    journal = "Front. Phys. (Beijing)",
    volume = "19",
    number = "1",
    pages = "14701",
    year = "2024"
}

@article{PhysRevLett.122.042001,
  title = {Observation of Transverse $\mathrm{\ensuremath{\Lambda}}/\overline{\mathrm{\ensuremath{\Lambda}}}$ Hyperon Polarization in ${e}^{+}{e}^{\ensuremath{-}}$ Annihilation at Belle},
  author = {Guan, Y. and others},
  collaboration = {Belle Collaboration},
  journal = {Phys. Rev. Lett.},
  volume = {122},
  issue = {4},
  pages = {042001},
  numpages = {8},
  year = {2019},
  month = {Jan},
  publisher = {American Physical Society},
  doi = {10.1103/PhysRevLett.122.042001},
  url = {https://link.aps.org/doi/10.1103/PhysRevLett.122.042001}
}

@article{PhysRevLett.36.1113,
  title = "{${\ensuremath{\Lambda}}^{0}$ Hyperon Polarization in Inclusive Production by 300-GeV Protons on Beryllium}",
  author = {Bunce, G. and others},
  journal = {Phys. Rev. Lett.},
  volume = {36},
  issue = {19},
  pages = {1113--1116},
  numpages = {0},
  year = {1976},
  month = {May},
  publisher = {American Physical Society},
  doi = {10.1103/PhysRevLett.36.1113},
  url = {https://link.aps.org/doi/10.1103/PhysRevLett.36.1113}
}

@article{PhysRevLett.67.1193,
  title = "{Polarization of ${\mathrm{\ensuremath{\Xi}}}^{\mathrm{\ensuremath{-}}}$ hyperons produced by 800-GeV protons}",
  author = {Duryea, J. and others},
  journal = {Phys. Rev. Lett.},
  volume = {67},
  issue = {10},
  pages = {1193--1196},
  numpages = {0},
  year = {1991},
  month = {Sep},
  publisher = {American Physical Society},
  doi = {10.1103/PhysRevLett.67.1193},
  url = {https://link.aps.org/doi/10.1103/PhysRevLett.67.1193}
}

@article{ABLIKIM2010345,
title = {Design and construction of the BESIII detector},
journal = {Nucl. Instr. Meth. A},
volume = {614},
number = {3},
pages = {345-399},
year = {2010},
issn = {0168-9002},
doi = {https://doi.org/10.1016/j.nima.2009.12.050},
url = {https://www.sciencedirect.com/science/article/pii/S0168900209023870},
author = {M. Ablikim and others}
}

@article{PhysRevLett.90.142301,
  title = {Measurement of Spin-Correlation Parameters ${A}_{NN}$, ${A}_{SS}$, and ${A}_{SL}$ at 2.1 GeV in Proton-Proton Elastic Scattering},
  author = {Bauer, F. and others},
  collaboration = {EDDA Collaboration},
  journal = {Phys. Rev. Lett.},
  volume = {90},
  issue = {14},
  pages = {142301},
  numpages = {4},
  year = {2003},
  month = {Apr},
  publisher = {American Physical Society},
  doi = {10.1103/PhysRevLett.90.142301},
  url = {https://link.aps.org/doi/10.1103/PhysRevLett.90.142301}
}

@article{Bystricky:1976jr,
    author = "Bystricky, J. and Lehar, F. and Winternitz, P.",
    title = "{Formalism of Nucleon-Nucleon Elastic Scattering Experiments}",
    reportNumber = "SACLAY-DPHPE-76-12",
    doi = "10.1051/jphys:019780039010100",
    journal = "J. Phys. (France)",
    volume = "39",
    pages = "1",
    year = "1978"
}

@article{Anderle:2021wcy,
    author = "Anderle, Daniele P. and others",
    title = "{Electron-ion collider in China}",
    eprint = "2102.09222",
    archivePrefix = "arXiv",
    primaryClass = "nucl-ex",
    reportNumber = "Frontiers of Physics, Volume 16 Issue (6):64701, 2021",
    doi = "10.1007/s11467-021-1062-0",
    journal = "Front. Phys. (Beijing)",
    volume = "16",
    number = "6",
    pages = "64701",
    year = "2021"
}

@article{AbdulKhalek:2021gbh,
    author = "Abdul Khalek, R. and others",
    title = "{Science Requirements and Detector Concepts for the Electron-Ion Collider}: {EIC Yellow Report}",
    eprint = "2103.05419",
    archivePrefix = "arXiv",
    primaryClass = "physics.ins-det",
    reportNumber = "BNL-220990-2021-FORE, JLAB-PHY-21-3198, LA-UR-21-20953",
    doi = "10.1016/j.nuclphysa.2022.122447",
    journal = "Nucl. Phys. A",
    volume = "1026",
    pages = "122447",
    year = "2022"
}

@article{PhysRev.148.1289,
  title = "{Polarization Parameter in $p\ensuremath{-}p$ Scattering from 328 to 736 MeV}",
  author = {Betz, F. and others},
  journal = {Phys. Rev.},
  volume = {148},
  issue = {4},
  pages = {1289--1296},
  numpages = {0},
  year = {1966},
  month = {Aug},
  publisher = {American Physical Society},
  doi = {10.1103/PhysRev.148.1289},
  url = {https://link.aps.org/doi/10.1103/PhysRev.148.1289}
}

@article{ALBROW1970445,
title = "{Polarization in elastic proton-proton scattering between 0.86 and 2.74 GeV/c}",
journal = {Nucl. Phys. B},
volume = {23},
number = {3},
pages = {445-465},
year = {1970},
issn = {0550-3213},
doi = {https://doi.org/10.1016/0550-3213(70)90296-8},
url = {https://www.sciencedirect.com/science/article/pii/0550321370902968},
author = {M.G. Albrow and others}
}

@article{PhysRevD.40.35,
  title = "{Elastic ${p}_{\ensuremath{\uparrow}}{p}_{\ensuremath{\uparrow}}$ scattering between 240 and 470 MeV}",
  author = {Onel, Y. and others},
  journal = {Phys. Rev. D},
  volume = {40},
  issue = {1},
  pages = {35--43},
  numpages = {0},
  year = {1989},
  month = {Jul},
  publisher = {American Physical Society},
  doi = {10.1103/PhysRevD.40.35},
  url = {https://link.aps.org/doi/10.1103/PhysRevD.40.35}
}

@article{PhysRevC.24.1778,
  title = "{$\mathrm{pp}$ elastic analyzing power from 318 to 800 MeV}",
  author = {McNaughton, M. W. and Chamberlin, E. P.},
  journal = {Phys. Rev. C},
  volume = {24},
  issue = {4},
  pages = {1778--1781},
  numpages = {0},
  year = {1981},
  month = {Oct},
  publisher = {American Physical Society},
  doi = {10.1103/PhysRevC.24.1778},
  url = {https://link.aps.org/doi/10.1103/PhysRevC.24.1778}
}

@article{vonPrzewoski:1998ye,
    author = "von Przewoski, B. and others",
    title = "{Proton proton analyzing power and spin correlation measurements between 250~MeV and 450~MeV at $7^\circ < \theta_{c.m.}< 90^\circ$  with an internal target in a storage ring}",
    doi = "10.1103/PhysRevC.58.1897",
    journal = "Phys. Rev. C",
    volume = "58",
    pages = "1897--1912",
    year = "1998"
}

@article{PhysRevLett.41.384,
  title = "{Polarization Analyzing Power ${A}_{y}(\ensuremath{\theta})$ in $\mathrm{pp}$ Elastic Scattering at 643, 787, and 796 MeV}",
  author = {Bevington, P. R. and others},
  journal = {Phys. Rev. Lett.},
  volume = {41},
  issue = {6},
  pages = {384--387},
  numpages = {0},
  year = {1978},
  month = {Aug},
  publisher = {American Physical Society},
  doi = {10.1103/PhysRevLett.41.384},
  url = {https://link.aps.org/doi/10.1103/PhysRevLett.41.384}
}

@article{PhysRev.105.288,
  title = "{Experiments with 315-Mev Po\-lari\-zed Protons: Proton-Proton and Proton-Neutron Scattering}",
  author = {Chamberlain, O. and others},
  journal = {Phys. Rev.},
  volume = {105},
  issue = {1},
  pages = {288--301},
  numpages = {0},
  year = {1957},
  month = {Jan},
  publisher = {American Physical Society},
  doi = {10.1103/PhysRev.105.288},
  url = {https://link.aps.org/doi/10.1103/PhysRev.105.288}
}

@article{GREENIAUS1979308,
title = "{Measurements of p-p and p-$^4$He ana\-ly\-zing powers at medium energies}",
journal = {Nucl. Phy. A},
volume = {322},
number = {2},
pages = {308-328},
year = {1979},
issn = {0375-9474},
doi = {https://doi.org/10.1016/0375-9474(79)90428-7},
url = {https://www.sciencedirect.com/science/article/pii/0375947479904287},
author = {L.G. Greeniaus and others},
}

@article{PhysRev.163.1470,
  title = "{Nucleon-Nucleon Polarization between 300 and 700 MeV}",
  author = {Cheng, David and others},
  journal = {Phys. Rev.},
  volume = {163},
  issue = {5},
  pages = {1470--1478},
  numpages = {0},
  year = {1967},
  month = {Nov},
  publisher = {American Physical Society},
  doi = {10.1103/PhysRev.163.1470},
  url = {https://link.aps.org/doi/10.1103/PhysRev.163.1470}
}

@article{PhysRev.95.1348,
  title = "{Small-Angle $p\ensuremath{-}p$ Cross Sections and Polarization at 300 MeV}",
  author = {Chamberlain, O. and others},
  journal = {Phys. Rev.},
  volume = {95},
  issue = {5},
  pages = {1348--1349},
  numpages = {0},
  year = {1954},
  month = {Sep},
  publisher = {American Physical Society},
  doi = {10.1103/PhysRev.95.1348},
  url = {https://link.aps.org/doi/10.1103/PhysRev.95.1348}
}

@article{PhysRevD.21.580,
  title = "{Measurement of the spin-dependent parameters $D$, $R$, $A$, and $P$ for small-angle $p\ensuremath{-}p$ elastic scattering between 300 and 600 MeV}",
  author = {Besset, D. and others},
  journal = {Phys. Rev. D},
  volume = {21},
  issue = {3},
  pages = {580--598},
  numpages = {0},
  year = {1980},
  month = {Feb},
  publisher = {American Physical Society},
  doi = {10.1103/PhysRevD.21.580},
  url = {https://link.aps.org/doi/10.1103/PhysRevD.21.580}
}

@article{ONUKI202278,
title = "{Belle II status and prospect}",
journal = {Nuclear and Particle Physics Proceedings},
volume = {318-323},
pages = {78-84},
year = {2022},
note = {QCD 21 is the 24th International Conference on Quantum Chromodynamics},
issn = {2405-6014},
doi = {https://doi.org/10.1016/j.nuclphysbps.2022.09.017},
url = {https://www.sciencedirect.com/science/article/pii/S2405601422000177},
author = {Yoshiyuki Onuki}
}

@article{Hayrapetyan_2024,
doi = {10.1088/1748-0221/19/05/P05064},
url = {https://dx.doi.org/10.1088/1748-0221/19/05/P05064},
year = {2024},
month = {may},
publisher = {IOP Publishing},
volume = {19},
number = {05},
pages = {P05064},
author = {Hayrapetyan, A. and others},
collaboration = {CMS},
title = {Development of the CMS detector for the CERN LHC Run 3},
journal = {Journal of Instrumentation}
}

@article{Aad_2024,
doi = {10.1088/1748-0221/19/05/P05063},
url = {https://dx.doi.org/10.1088/1748-0221/19/05/P05063},
year = {2024},
month = {may},
publisher = {IOP Publishing},
volume = {19},
number = {05},
pages = {P05063},
author = {Aad, G. and others},
collaboration = {ATLAS},
title = {The ATLAS experiment at the CERN Large Hadron Collider: a description of the detector configuration for Run 3},
journal = {Journal of Instrumentation}
}

@article{c642-1lzb,
  title = {How to determine nucleon polarization at existing collider experiments?},
  author = {Liang, Yu-Tie and Lv, Xiao-Rong and Kupsc, Andrzej and Gou, Boxing and Li, Hai-Bo},
  journal = {Phys. Rev. D},
  volume = {112},
  issue = {3},
  pages = {L031502},
  numpages = {6},
  year = {2025},
  month = {Aug},
  publisher = {American Physical Society},
  doi = {10.1103/c642-1lzb},
  url = {https://link.aps.org/doi/10.1103/c642-1lzb}
}

@article{CEPC,
    author = {Dong, Mingyi and others},
    title = "{CEPC Conceptual Design Report}",
    eprint = "1811.10545",
    archivePrefix = "arXiv",
    primaryClass = "hep-ex",
    month = "11",
    year = "2018"
}

@article{PhysRevC.41.1651,
  title = "{Cross sections, analyzing powers, and spin-rotation-depolarization observables for 500 MeV proton elastic scattering from $^{12}\mathrm{C}$ and $^{13}\mathrm{C}$}",
  author = {Hoffmann, G. W. and others},
  journal = {Phys. Rev. C},
  volume = {41},
  issue = {4},
  pages = {1651--1655},
  numpages = {0},
  year = {1990},
  month = {Apr},
  publisher = {American Physical Society},
  doi = {10.1103/PhysRevC.41.1651},
  url = {https://link.aps.org/doi/10.1103/PhysRevC.41.1651}
}

@article{10.1093/ptep/ptx116,
    author = {Kaki, Kaori},
    title = "{Reaction cross sections of proton scattering from carbon isotopes (A=8-22) by means of the relativistic impulse approximation}",
    journal = {Prog. Theo. Exp. Phys.},
    volume = {2017},
    number = {9},
    pages = {093D01},
    year = {2017},
    month = {09},
    issn = {2050-3911},
    doi = {10.1093/ptep/ptx116},
    url = {https://doi.org/10.1093/ptep/ptx116}
}

@article{AGOSTINELLI2003250,
title = "{Geant4—a simulation toolkit}",
journal = {Nucl. Instr. Meth. A},
volume = {506},
number = {3},
pages = {250-303},
year = {2003},
issn = {0168-9002},
doi = {https://doi.org/10.1016/S0168-9002(03)01368-8},
url = {https://www.sciencedirect.com/science/article/pii/S0168900203013688},
author = {S. Agostinelli and others}
}

@article{STAR:2017ckg,
    author = "Adamczyk, L. and others",
    collaboration = "STAR Collaboration",
    title = "{Global $\Lambda$ hyperon polarization in nuclear collisions: evidence for the most vortical fluid}",
    eprint = "1701.06657",
    archivePrefix = "arXiv",
    primaryClass = "nucl-ex",
    doi = "10.1038/nature23004",
    journal = "Nature",
    volume = "548",
    pages = "62--65",
    year = "2017"
}

@article{STAR:2022fan,
    author = "Abdallah, M. S. and others",
    collaboration = "STAR Collaboration",
    title = "{Pattern of global spin alignment of {\ensuremath{\phi}} and K$^{*0}$ mesons in heavy-ion collisions}",
    eprint = "2204.02302",
    archivePrefix = "arXiv",
    primaryClass = "hep-ph",
    doi = "10.1038/s41586-022-05557-5",
    journal = "Nature",
    volume = "614",
    number = "7947",
    pages = "244--248",
    year = "2023"
}

@article{Chen:2024aom,
    author = "Chen, Jinhui and others",
    title = "{Properties of the QCD matter: review of selected results from the relativistic heavy ion collider beam energy scan (RHIC BES) program}",
    eprint = "2407.02935",
    archivePrefix = "arXiv",
    primaryClass = "nucl-ex",
    doi = "10.1007/s41365-024-01591-2",
    journal = "Nucl. Sci. Tech.",
    volume = "35",
    number = "12",
    pages = "214",
    year = "2024"
}

@article{SAID_2000,
  title = "{Nucleon-nucleon elastic scattering to 3 GeV}",
  author = {Arndt, Richard A. and Strakovsky, Igor I. and Workman, Ron L.},
  journal = {Phys. Rev. C},
  volume = {62},
  issue = {3},
  pages = {034005},
  numpages = {16},
  year = {2000},
  month = {Aug},
  publisher = {American Physical Society},
  doi = {10.1103/PhysRevC.62.034005},
  url = {https://link.aps.org/doi/10.1103/PhysRevC.62.034005}
}

@article{SAID_2007,
  title = "{Updated analysis of $NN$ elastic scattering to 3 GeV}",
  author = {Arndt, R. A. and others},
  journal = {Phys. Rev. C},
  volume = {76},
  issue = {2},
  pages = {025209},
  numpages = {10},
  year = {2007},
  month = {Aug},
  publisher = {American Physical Society},
  doi = {10.1103/PhysRevC.76.025209},
  url = {https://link.aps.org/doi/10.1103/PhysRevC.76.025209}
}

@article{MAID_1998,
title = "{A unitary isobar model for pion photo- and electroproduction on the proton up to 1 GeV}",
journal = {Nucl. Phys. A},
volume = {645},
number = {1},
pages = {145-174},
year = {1999},
issn = {0375-9474},
doi = {https://doi.org/10.1016/S0375-9474(98)00572-7},
url = {https://www.sciencedirect.com/science/article/pii/S0375947498005727},
author = {D. Drechsel and O. Hanstein and S.S. Kamalov and L. Tiator}
}

@article{MAID_2007,
   author       = "D. Drechsel and others",
   title = "{Unitary isobar model - MAID2007}",
   year         = "2007",
   journal      = "Eur. Phys. J. A",
   volume       = "34",
   pages        = "69",
   doi = {https://doi.org/10.1140/epja/i2007-10490-6}
}

@misc{NNOnline,
   author       = "NN-OnLine",
   note         = "{https://nn-online.org}"
}

@misc{SAIDURL,
   note         = "{http://gwdac.phys.gwu.edu}"
}

@misc{MAIDURL,
   note         = "{https://maid.kph.uni-mainz.de}"
}

@article{PhysRevLett.94.102301,
  title = "{Globally Polarized Quark-Gluon Plasma in Noncentral $A+A$ Collisions}",
  author = {Liang, Zuo-Tang and Wang, Xin-Nian},
  journal = {Phys. Rev. Lett.},
  volume = {94},
  issue = {10},
  pages = {102301},
  numpages = {4},
  year = {2005},
  month = {Mar},
  publisher = {American Physical Society},
  doi = {10.1103/PhysRevLett.94.102301},
  url = {https://link.aps.org/doi/10.1103/PhysRevLett.94.102301}
}

@article{PhysRevC.77.044902,
  title = "{Global quark polarization in noncentral $A+A$ collisions}",
  author = {Gao, Jian-Hua and others},
  journal = {Phys. Rev. C},
  volume = {77},
  issue = {4},
  pages = {044902},
  numpages = {13},
  year = {2008},
  month = {Apr},
  publisher = {American Physical Society},
  doi = {10.1103/PhysRevC.77.044902},
  url = {https://link.aps.org/doi/10.1103/PhysRevC.77.044902}
}

@article{Liang:2004xn,
    author = "Liang, Zuo-Tang and Wang, Xin-Nian",
    title = "{Spin alignment of vector mesons in non-central A+A collisions}",
    eprint = "nucl-th/0411101",
    archivePrefix = "arXiv",
    reportNumber = "LBNL-56659",
    doi = "10.1016/j.physletb.2005.09.060",
    journal = "Phys. Lett. B",
    volume = "629",
    pages = "20--26",
    year = "2005"
}

@article{Liu:2000fi,
    author = "Liu, Chun-Xiu and Liang, Zuo-Tang",
    title = "{Spin structure and longitudinal polarization of hyperon in e+ e- annihilation at high-energies}",
    eprint = "hep-ph/0005172",
    archivePrefix = "arXiv",
    doi = "10.1103/PhysRevD.62.094001",
    journal = "Phys. Rev. D",
    volume = "62",
    pages = "094001",
    year = "2000"
}

@article{Herui-NST,
    author = "He, Rui and others",
    title = "{Advances in nuclear detection and readout techniques}",
    doi = "10.1007/s41365-023-01359-0",
    journal = "Nucl. Sci. Tech.",
    volume = "34",
    pages = "205",
    year = "2023"
}

@article{Liukang-NST,
    author = "Liu, Kang and others",
    title = "{Performance of AC-LGAD strip sensors designed for the DarkSHINE experiment}",
    doi = "10.1007/s41365-024-01575-2",
    journal = "Nucl. Sci. Tech.",
    volume = "35",
    pages = "201",
    year = "2024"
}

@article{ECal-NST,
    author = "Yu, Xiao-Zhou and others",
    title = "{Production and test of sPHENIX W/SciFiber electromagnetic calorimeter blocks in China}",
    doi = "10.1007/s41365-024-01517-y",
    journal = "Nucl. Sci. Tech.",
    volume = "35",
    pages = "145",
    year = "2024"
}

@article{Zhangjinlong-NST,
    author = "Ji, Zhaohuizi and others",
    title = "{Lambda polarization at the Electron-ion collider in China}",
    doi = "10.1007/s41365-023-01317-w",
    journal = "Nucl. Sci. Tech.",
    volume = "34",
    pages = "155",
    year = "2023"
}

@article{CHEN2023874,
title = "{Global spin alignment of vector mesons and strong force fields in heavy-ion collisions}",
journal = {Sci. Bull.},
volume = {68},
number = {9},
pages = {874-877},
year = {2023},
issn = {2095-9273},
doi = {https://doi.org/10.1016/j.scib.2023.04.001},
url = {https://www.sciencedirect.com/science/article/pii/S2095927323002335},
author = {Jinhui Chen and Zuo-Tang Liang and Yu-Gang Ma and Qun Wang}
}

@article{LU1995419,
title = "{The strange quark spin of the proton in semi-inclusive $\Lambda$ leptoproduction}",
journal = {Phys. Lett. B},
volume = {357},
number = {3},
pages = {419-422},
year = {1995},
issn = {0370-2693},
doi = {https://doi.org/10.1016/0370-2693(95)00927-D},
url = {https://www.sciencedirect.com/science/article/pii/037026939500927D},
author = {Wei Lu and Bo-Qiang Ma}
}

@article{DuanFF-NST,
    author = "Duan, Fang-Fang and others",
    title = "{Silicon detector array for radioactive beam experiments at HIRFL-RIBLL}",
    doi = "10.1007/s41365-018-0499-5",
    journal = "Nucl. Sci. Tech.",
    volume = "29",
    pages = "165",
    year = "2018"
}

@article{Wangshen-NST,
    author = "Wang, Shen and others",
    title = "{Design and testing of a miniature silicon strip detector}",
    doi = "10.1007/s41365-019-0714-z",
    journal = "Nucl. Sci. Tech.",
    volume = "31",
    pages = "7",
    year = "2020"
}

@article{Cao-JINST,
    author = "Cao, B and others",
    title = "{Study of MIPs effects on a MAPS for electron ion collider in China}",
    doi = "10.1088/1748-0221/18/05/C05016",
    journal = "J. Instrum.",
    volume = "18",
    pages = "C05016",
    year = "2023"
}

@article{ZhaoC-JINST,
    author = "Zhao, C and others",
    title = "{Study of the charge sensing node in the MAPS for therapeutic carbon ion beams}",
    doi = "10.1088/1748-0221/14/05/C05006",
    journal = "J. Instrum.",
    volume = "14",
    pages = "C05006",
    year = "2019"
}

@article{Malong-NST,
    author = "Ma, Long and others",
    title = "{Alignment calibration and performance study of the STAR PXL detector}",
    doi = "10.1007/s41365-016-0177-4",
    journal = "Nucl. Sci. Tech.",
    volume = "28",
    pages = "25",
    year = "2017"
}

@article{Xu:2005ru,
    author = "Xu, Qing-Hua and Liang, Zuo-Tang and Sichtermann, Ernst",
    title = "{Anti-lambda polarization in high energy pp collisions with polarized beam}",
    eprint = "hep-ph/0511061",
    archivePrefix = "arXiv",
    reportNumber = "LBNL-59058",
    doi = "10.1103/PhysRevD.73.077503",
    journal = "Phys. Rev. D",
    volume = "73",
    pages = "077503",
    year = "2006"
}

@article{Xu:2004es,
    author = "Xu, Qing-Hua and Liang, Zuo-Tang",
    title = "{Probing gluon helicity distribution and quark transversity through hyperon polarization in singly polarized pp collisions}",
    eprint = "hep-ph/0406119",
    archivePrefix = "arXiv",
    doi = "10.1103/PhysRevD.70.034015",
    journal = "Phys. Rev. D",
    volume = "70",
    pages = "034015",
    year = "2004"
}

@article{Zhou:2008fb,
    author = "Zhou, Jian and Yuan, Feng and Liang, Zuo-Tang",
    title = "{Hyperon Polarization in Unpolarized Scattering Processes}",
    eprint = "0808.3629",
    archivePrefix = "arXiv",
    primaryClass = "hep-ph",
    doi = "10.1103/PhysRevD.78.114008",
    journal = "Phys. Rev. D",
    volume = "78",
    pages = "114008",
    year = "2008"
}

@article{PhysRevLett.134.022301,
  title = "{Deciphering Hypertriton and Antihypertriton Spins from Their Global Polarizations in Heavy-Ion Collisions}",
  author = {Sun, Kai-Jia and Liu, Dai-Neng and Zheng, Yun-Peng and Chen, Jin-Hui and Ko, Che Ming and Ma, Yu-Gang},
  journal = {Phys. Rev. Lett.},
  volume = {134},
  issue = {2},
  pages = {022301},
  numpages = {6},
  year = {2025},
  month = {Jan},
  publisher = {American Physical Society},
  doi = {10.1103/PhysRevLett.134.022301},
  url = {https://link.aps.org/doi/10.1103/PhysRevLett.134.022301}
}

\end{document}